\documentclass[final,5p,times,twocolumn,numbers,sort,compress]{elsarticle}
\usepackage{hyperref}

\hypersetup{
	      pdfstartview=FitH,
            CJKbookmarks=true,
            bookmarksnumbered=true,
            bookmarksopen=true,
            colorlinks, 
            pdfborder=001,   
            linkcolor=black,
            anchorcolor=black,
            linkcolor=black,
            citecolor=black
}

\usepackage{amssymb}
\usepackage{lipsum}
\usepackage{amsmath}

\usepackage{graphicx}
\graphicspath{{./Figures/}}
\usepackage{xcolor}
\usepackage{ulem}
\usepackage{dcolumn}
\usepackage{bm}

\usepackage[caption=false]{subfig}
\usepackage{siunitx}
\usepackage[capitalise]{cleveref}

\journal{Physics Letters B}

\begin{document}

\begin{frontmatter}

\title{Global analysis of Sivers and Collins asymmetries within the TMD factorization}

\author[IMP,LZU,UCAS]{Chunhua Zeng}
\ead{zengchunhua@impcas.ac.cn}

\author[NJNU]{Hongxin Dong}
\ead{hxdong@nnu.edu.cn}

\author[SDU,SCNT]{Tianbo Liu}
\ead{liutb@sdu.edu.cn}

\author[IMP]{Peng Sun}
\ead{pengsun@impcas.ac.cn}

\author[IMP,UCAS,SCNT,CCNU]{Yuxiang Zhao}
\ead{yxzhao@impcas.ac.cn}

\address[IMP]{Institute of Modern Physics, Chinese Academy of Sciences, Lanzhou, Gansu 730000, China}
\address[LZU]{Lanzhou University, Lanzhou, Gansu 730000, China}
\address[UCAS]{University of Chinese Academy of Sciences, Beijing 100049, China}
\address[NJNU]{Department of Physics and Institute of Theoretical Physics,
Nanjing Normal University, Nanjing, Jiangsu 210023, China}
\address[SDU]{Key Laboratory of Particle Physics and Particle Irradiation (MOE), Institute of Frontier and Interdisciplinary Science, Shandong University, Qingdao, Shandong 266237, China}
\address[SCNT]{Southern Center for Nuclear-Science Theory (SCNT), Institute of Modern Physics, Chinese Academy of Sciences, Huizhou, Guangdong 516000, China}
\address[CCNU]{Key Laboratory of Quark and Lepton Physics (MOE) and Institute of Particle Physics, Central China Normal University, Wuhan 430079, China}

\author{\\ (Transverse Nucleon Tomography Collaboration)}

\begin{abstract}

We present a global analysis of Sivers functions, transversity distribution functions, and Collins fragmentation functions within the transverse momentum dependent factorization. This analysis encompasses the latest data from semi-inclusive deep inelastic scattering, Drell-Yan, and $W^{\pm}/Z$-boson production processes as recently reported by the COMPASS and STAR Collaborations.
Upon integrating the new data into our fitting, 
the precision of the extracted $d$ and $\bar{d}$ quark Sivers and transversity distributions, as well as the tensor charge, is notably improved.

\end{abstract}


\end{frontmatter}




\section{Introduction}
\label{introduction}
The exploration of nucleon structure is one of the frontiers of modern hadron physics, providing critical insights into the fundamental nature of matter.
The underline theory is quantum chromodynamics (QCD).
The color confinement, a pivotal feature of the QCD, poses a significant barrier to the direct observation of isolated quarks and gluons, as they are perpetually confined within nucleons, or more broadly hadrons. Moreover, QCD has another essential feature called asymptotic freedom. This property indicates that the strong interaction becomes weakly coupled at high-energy scales, allowing for the formulation of the theoretical framework known as QCD factorization. According to the factorization, one can extract partonic structures of the nucleon through various experimental measurements, in which the cross section is approximated as a convolution of perturbatively calculable short-distance scattering of partons and universal long-distance functions~\cite{Collins:1984kg, Collins:1989gx}. The measurement of the partonic structure of the nucleon was pioneered by SLAC~\cite{Bloom:1969kc} via inclusive lepton-nucleon deep inelastic scattering (DIS) process, in which only the scattered lepton was detected. In the DIS process, the cross section can be expressed as the convolution of lepton-parton hard scattering cross section and the collinear parton distribution function (PDF) $f_1^q(x, \mu)$, which denotes the probability density of finding a parton of flavor $q$ carrying a certain longitudinal momentum fraction $x$ of the nucleon, with $\mu$ as the factorization scale.
 
Developing the theoretical framework to the transverse momentum-dependent (TMD) factorization, one can extract the three-dimensional motion of quarks and gluons within the colliding nucleon by identifying a final-state hadron with its transverse momentum much smaller than the hard scale. 
Taking into account the polarization of the nucleon and the parton, one can define eight TMD PDFs at the leading twist. These TMD PDFs describe the three-dimensional momentum of partons and their correlations with the spins of the nucleon and the quark. In this study, we focus on the transversity distribution function $h_1$
and the Sivers function $f_{1T}^{\perp}$.

The transversity distribution characterizes the net density of transversely polarized quarks within a transversely polarized proton. Its integral corresponds to the tensor charge, indicating the involvement of a tensor current in the interaction. As a chiral-odd quantity~\cite{Artru:1989zv, Jaffe:1991kp}, a practical way to access the transversity distribution is by coupling with another chiral-odd quantity, either a fragmentation function (FF) in semi-inclusive DIS (SIDIS) process~\cite{Collins:1992kk, Bacchetta:2008wb} or a distribution function in hadron-hadron collisions~\cite{Cortes:1991ja, Anselmino:2004ki,Efremov:2004qs,Pasquini:2006iv}. The Sivers function is often interpreted as the density of unpolarized quark within a transversely polarized nucleon. It was initially believed to vanish due to the time-reversal invariance of QCD~\cite{Collins:1992kk}, but this assertion applies specifically to transverse-momentum-integrated distribution functions. The inclusion of the nontrivial gauge link in the definition of gauge-invariant distribution functions challenges the validity of the original proof regarding distribution functions that depend on transverse momentum~\cite{Brodsky:2002cx,Collins:2002kn,Ji:2002aa,Belitsky:2002sm}. This gauge link provides the necessary phase for the interference related to T-oddness~\cite{Burkardt:2003yg}.

In recent studies, multiple analyses have been conducted to extract Sivers functions~\cite{Anselmino:2005ea,Collins:2005ie,Anselmino:2008sga,Bacchetta:2011gx,Sun:2013hua,Echevarria:2014xaa,Boglione:2018dqd,Echevarria:2020hpy,Bury:2020vhj,Bacchetta:2020gko} and TMD transversity distributions~\cite{Anselmino:2007fs,  Anselmino:2013vqa,   Anselmino:2015sxa,  Kang:2015msa, Lin:2017stx, DAlesio:2020vtw}.  Notably, references \cite{Cammarota:2020qcw, Gamberg:2022kdb, Boglione:2024dal} have extracted both the Sivers and transversity functions. It is worth noting that the extractions of TMDs in many early explorations~\cite{Anselmino:2005ea, Collins:2005ie, Anselmino:2008sga, Bacchetta:2011gx, Anselmino:2007fs,  Anselmino:2013vqa, Anselmino:2015sxa} and some recent studies~\cite{Boglione:2018dqd, Gamberg:2022kdb, Cammarota:2020qcw, Boglione:2024dal, Lin:2017stx, DAlesio:2020vtw} have not considered TMD energy evolution effect. The TMD evolution equations~\cite{Collins:2011zzd} were used in~\cite{Sun:2013hua, Echevarria:2014xaa, Echevarria:2020hpy, Bacchetta:2020gko, Bury:2020vhj} for the extraction of the Sivers functions and in~\cite{Kang:2015msa} for the extraction of the  transversity functions.

On the experimental aspect, many efforts for the SIDIS measurement of target transverse single
spin asymmetry (SSA) have been made during the last two decades by HERMES~\cite{HERMES:2020ifk}, COMPASS~\cite{COMPASS:2008isr,COMPASS:2014bze}, and Jefferson Lab (JLab)~\cite{JeffersonLabHallA:2011ayy,JeffersonLabHallA:2014yxb} for explorations of nucleon
TMD PDFs.
While providing very valuable information of nucleon three-dimensional spin structures, the uncertainties remain large in the extraction of TMDs when solely relying on these datasets~\cite{Zeng:2022lbo, Zeng:2023nnb}. The new measurements of the Collins and Sivers asymmetries, based on the data taken in 2022 by COMPASS using 160\,GeV muons on transversely polarized deuteron targets~\cite{COMPASS:2023vhr}, are outlined  on a high statistics, making it possible for a preciser determination of transversity and Sivers distributions.

In this study, we conduct a comprehensive analysis of the Collins and Sivers effects with the TMD evolution, incorporating the latest data from the COMPASS. For the Collins effect, we include measurements from semi-inclusive $ e^+e^- $ annihilation (SIA) data to simultaneously extract the Collins FFs, which serve as the chiral-odd partner to couple with the transversity distributions. Our analysis of the Sivers effect also encompasses the data from Drell-Yan (DY) and $W^{\pm}/Z$-boson production processes. The remaining paper is organized as follows. In Sec.~\ref{sec:data_set}, we briefly summarize the datasets of Sivers and Collins asymmetries.  In Sec.~\ref{sec:fit} we present the global analysis of world data and the fit results. A summary is provided in Sec.~\ref{sec:summary}.

\section{Data selection}\label{sec:data_set}
\begin{figure}[htp]
    \centering
    \includegraphics[width=0.95\columnwidth]{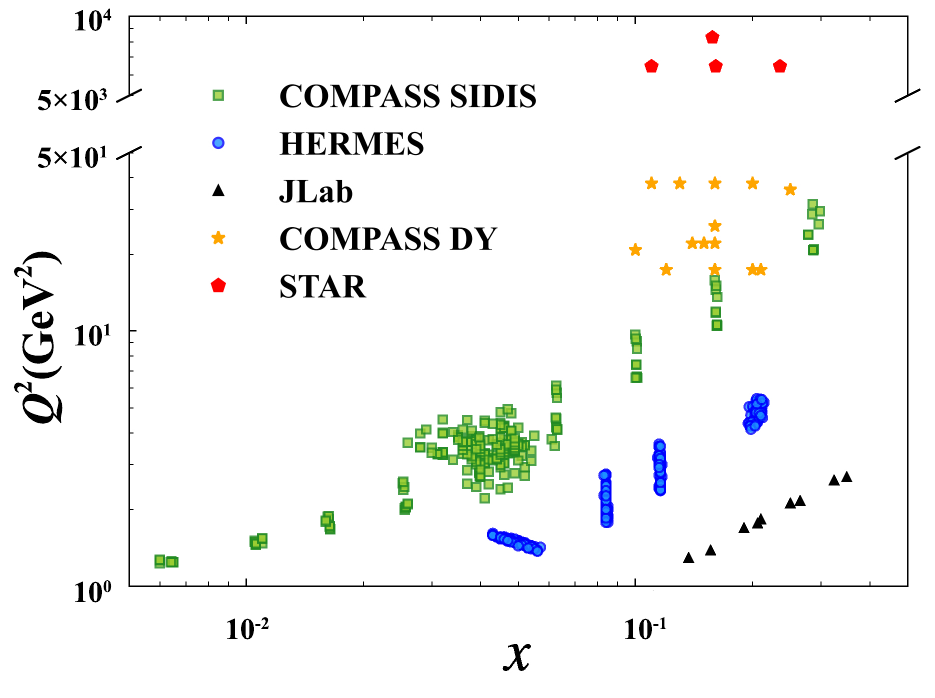}
    \caption{The kinematic distributions of the data for SIDIS,  Drell-Yan, and $W^{\pm}/Z$-boson production in the $x - Q^2$ plane.}
    \label{fig:world_data_kin}
\end{figure}

Our previous research ~\cite{Zeng:2022lbo, Zeng:2023nnb} involved an in-depth analysis of the Sivers and Collins asymmetries within the SIDIS process, drawing data from HERMES~\cite{HERMES:2020ifk}, COMPASS~\cite{COMPASS:2008isr, COMPASS:2014bze}, and JLab Hall A~\cite{JeffersonLabHallA:2011ayy, JeffersonLabHallA:2014yxb}. Recently, new data from COMPASS~\cite{COMPASS:2023vhr} collected from the 2022 run, featuring a transversely polarized deuteron target. These new data provide new insights into the Sivers and Collins asymmetries in the SIDIS process, based on a high-statistics measurement. The comprehensive global datasets on Sivers and Collins asymmetries, including the latest measurements from COMPASS~\cite{COMPASS:2023vhr} related to the SIDIS process, are summarized in Table~\ref{table:SIDIS_data}. For the purpose to extract TMDs, we focus on low transverse
momentum data with $\delta = |P_{h \perp}|/(zQ) < 1$, which aligns with the applicability of TMD factorization. Here $P_{h \perp}$ is the hadron transverse momentum in the virtual photon-nucleon frame, $z = P\cdot P_h / P\cdot q$, and $Q$ is the virtuality of the photon, serving as the hard scale. In addition to the original publications, the HERMES Collaboration recently presented re-analyzed data with three-dimensional binning in the kinematic variables $x$, $z$ 
and $P_{h\perp}$~\cite{HERMES:2020ifk}. Notably, within the region of the largest $P_{h\perp}$ bin, the small transverse
momentum expansion can not be used any more. To address this issue, we probably need to consider the impact of the corrections from
higher orders of $P_{h\perp}^2/Q^2$ and strong coupling constant $\alpha_s$. We will leave this issue of the analysis for future work. For the current study, we exclude the data from the largest $P_{h\perp}$ bin of HERMES~\cite{HERMES:2020ifk} in the fit. Furthermore, the presence of the Sivers function $f_{1T}^{\perp}$ not only induces the SSA in SIDIS process but also leads to corresponding SSA in Drell-Yan and the $W^{\pm}/Z$ production processes. These phenomena have been examined in measurements conducted by COMPASS ~\cite{COMPASS:2023vqt} and STAR~\cite{STAR:2015vmv, Collaboration:2023oml}. The datasets are listed in Table~\ref{table:DY_data}, with $\delta = q_T/Q < 1$ applied for TMD analysis, where $q_T$ is the transverse momentum of the vector boson. The kinematic distributions of the data for SIDIS,  Drell-Yan, and $W^{\pm}/Z$-boson production in the
$x - Q^2$ plane are shown in Fig.~\ref{fig:world_data_kin}. Each bin is plotted as
a point at the center kinematic values of the bin. 

For the selection of Collins asymmetry data within the SIA process, we keep the same criteria as in our previous work~\cite{Zeng:2023nnb}. The data from BELLE~\cite{Belle:2008fdv}, BABAR~\cite{BaBar:2013jdt,BaBar:2015mcn}, and BESIII~\cite{BESIII:2015fyw} are outlined in Table~\ref{table:SIA_data}, and $\delta < 1$ is applied similar to the SIDIS process.

\section{TMD extraction}\label{sec:fit}

We follow the theoretical formalism, usually referred to as the $\zeta$-prescription, which has been utilized in the analysis~\cite{Bury:2021sue} and our previous works~\cite{Zeng:2022lbo, Zeng:2023nnb} on Sivers and Collins asymmetries. In this study, we adopt a consistent parametrization for both Sivers and transversity distribution functions, $f^{\perp}_{1T}(x, b)$ and $h_1(x, b)$, of quark flavors $u$, $d$, $\bar{u}$, and $\bar{d}$. Detailed expressions of $f^{\perp}_{1T}(x, b)$,  $h_1(x, b)$, and  Collins FFs $H_1^{\perp}(z, b)$ to pions and kaons are provided in~\cite{Zeng:2023nnb}, where $b$ is the Fourier conjugate variable to the transverse
momentum of the parton.

Taking into account the correlation among experimental data points, one can express the  $\chi^2$ as
\begin{align}
    \chi^2=\sum_{i, j=1}^n(m_i-t_i)V_{ij}^{-1}(m_j-t_j),
\end{align}
where $m_i$ represents the central value of the $i$'th measurement, $t_i$  is corresponding theoretical value, and $V_{ij}$ is the covariance matrix.  The main source of correlated uncertainties in the SSA measurements arises from the target polarization and the dilution factors, which lead to an overall relative scale uncertainty $\sigma^{ cor.}$. In this scenario, the covariance matrix can be expressed as
\begin{align}
    V_{ij}=&\delta_{ij}(\sigma_i^{uncor.})^2[1+(\sigma^{ cor.})^2]+(\sigma^{ cor.})^2m_im_j,
\end{align}
where $\sigma_i^{uncor.}$ represents the uncorrelated uncertainty of the $i'$th measurement, which is evaluated via the quadratic sum of statistical and uncorrelated systematic uncertainties. To estimate the uncertainties of extracted TMD functions, we generate 1000 replicas of world data by randomly shifting the central values of the data points according to correlated and uncorrelated uncertainties. The details of this process are outlined in ~\ref{App:smear_replicas}. Then we perform a fit to each replica.
The values of $\chi^2/N$ for the fits corresponding to the SIDIS, DY, and SIA processes are listed in Tables~\ref{table:SIDIS_data} to~\ref{table:SIA_data}, respectively. Here, 
$N$ denotes the number of experimental data points. The comparisons between experimental data and the fit results are shown in supplemental material~\cite{Zeng:2024_sup_mat}.

\begin{figure*}[htp]
		\centering
		\includegraphics[width=0.4\textwidth]{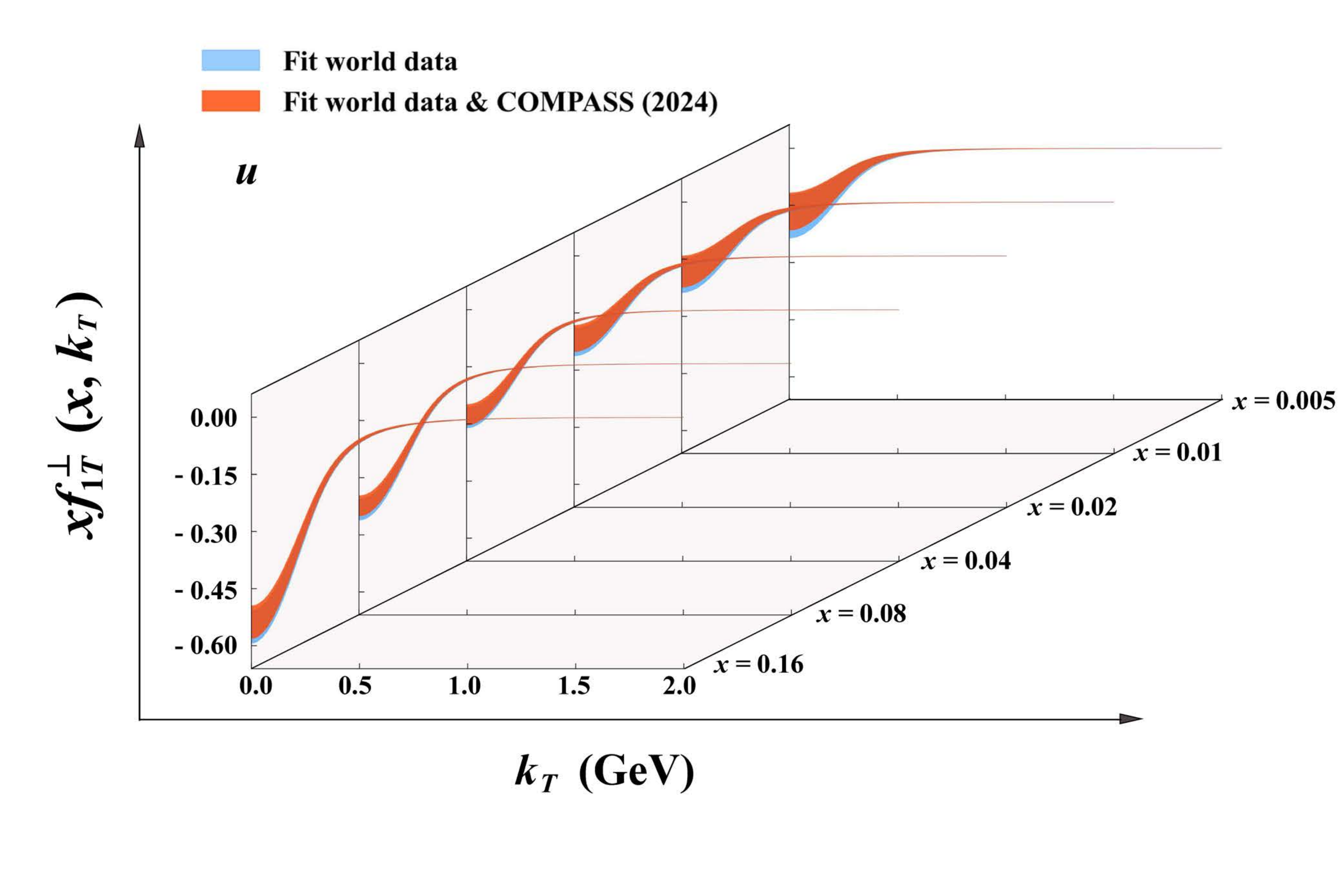}
		\includegraphics[width=0.4\textwidth]{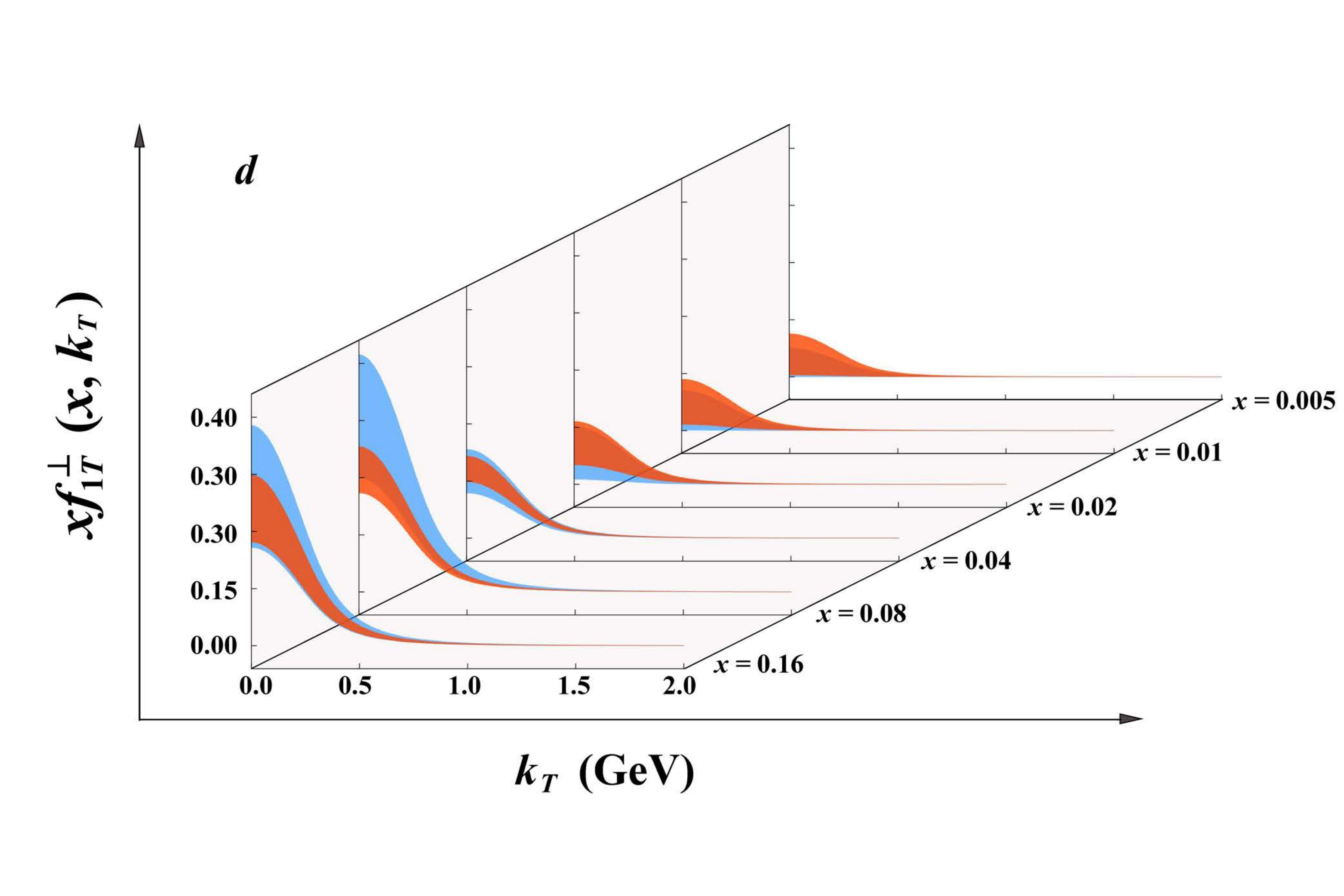}
		\includegraphics[width=0.4\textwidth]{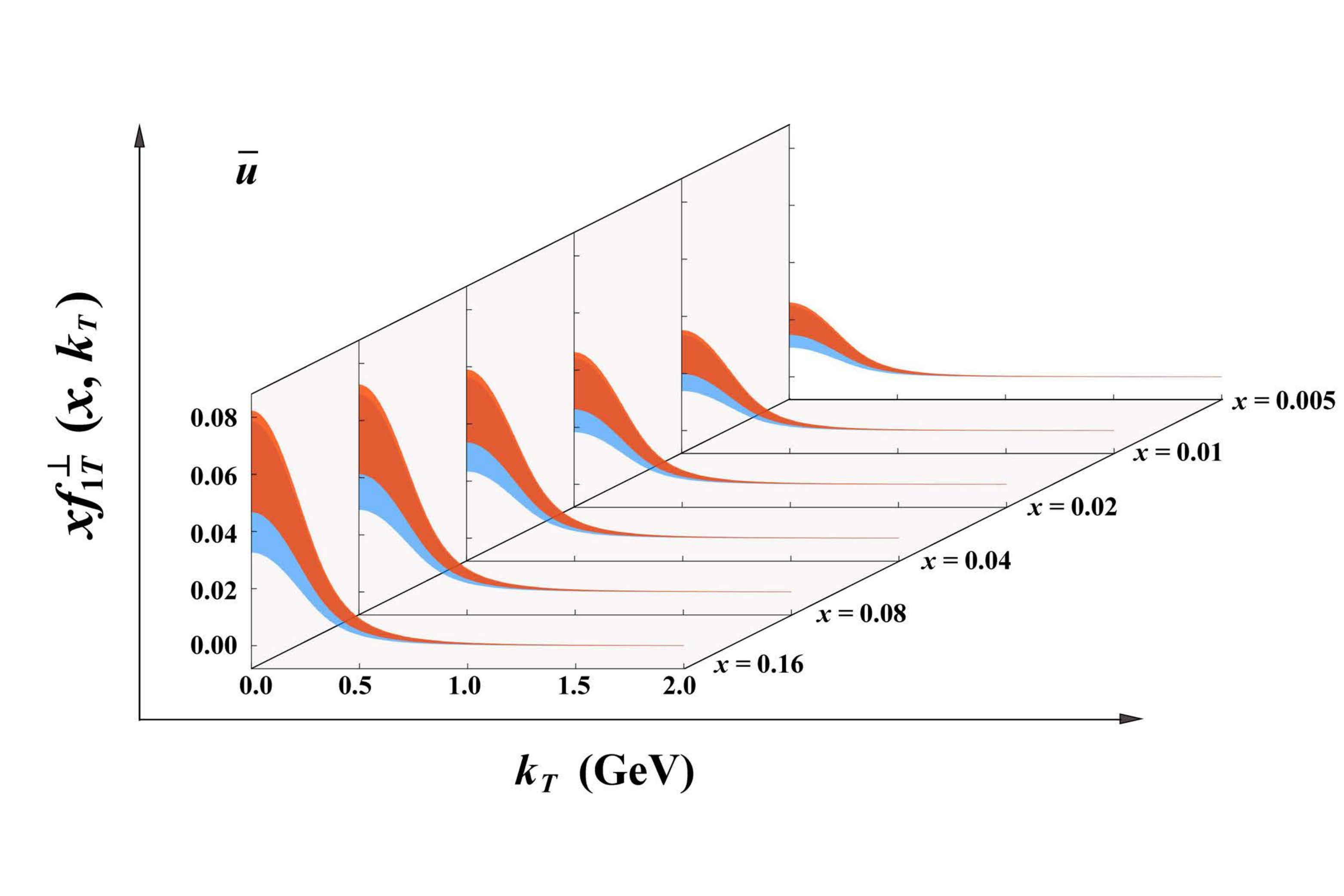}
		\includegraphics[width=0.4\textwidth]{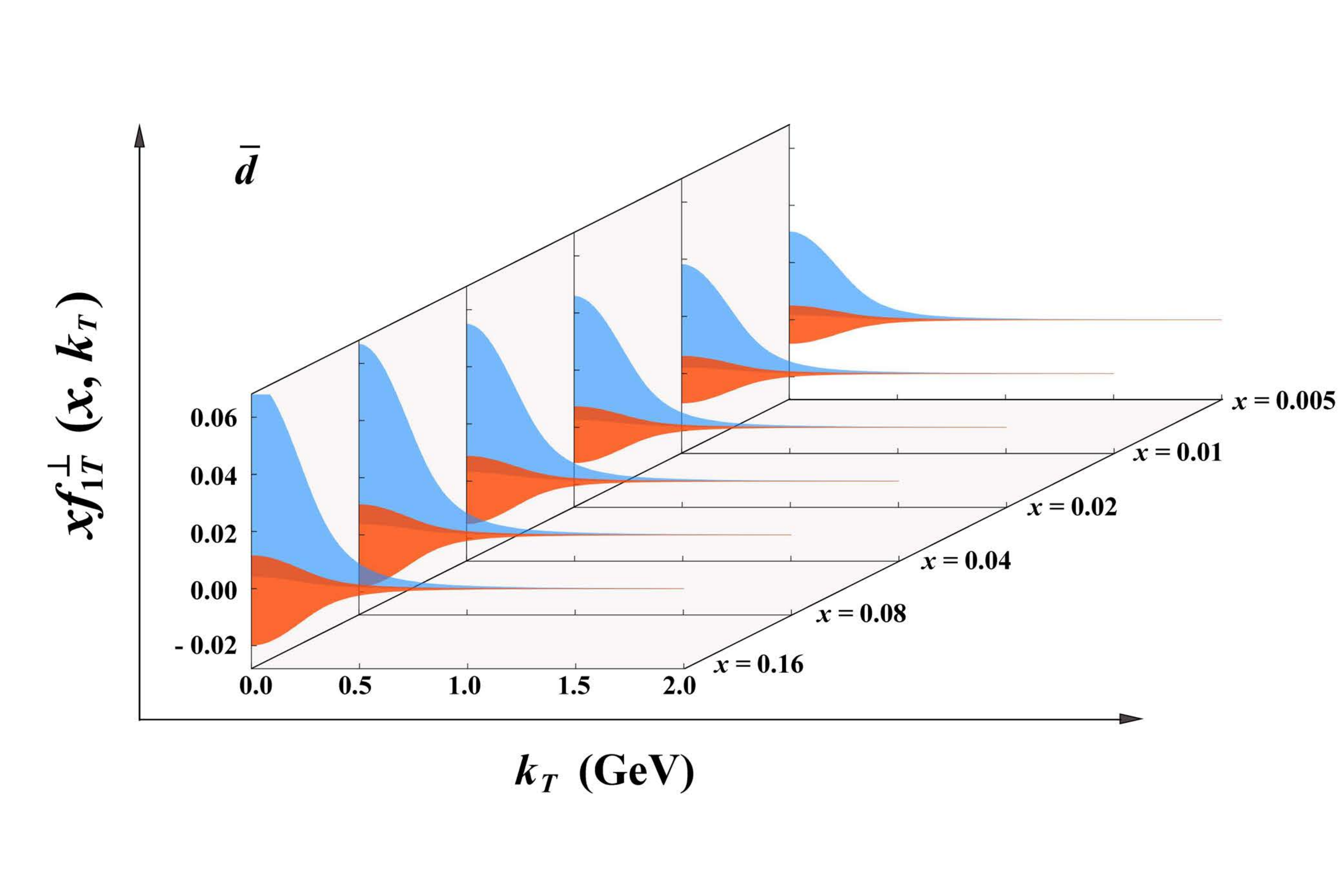}
  
		\caption{The transverse momentum distribution of the Sivers functions at different $x$ values and $Q=2\, \rm GeV$. The blue (red) bands represent the uncertainties of the fit to the world data without (with) the  new COMPASS data~\cite{COMPASS:2023vhr}.}
		\label{fig:xfTx_xslices}
\end{figure*}

\begin{figure*}[htp]
		\centering
		\includegraphics[width=0.4\textwidth]{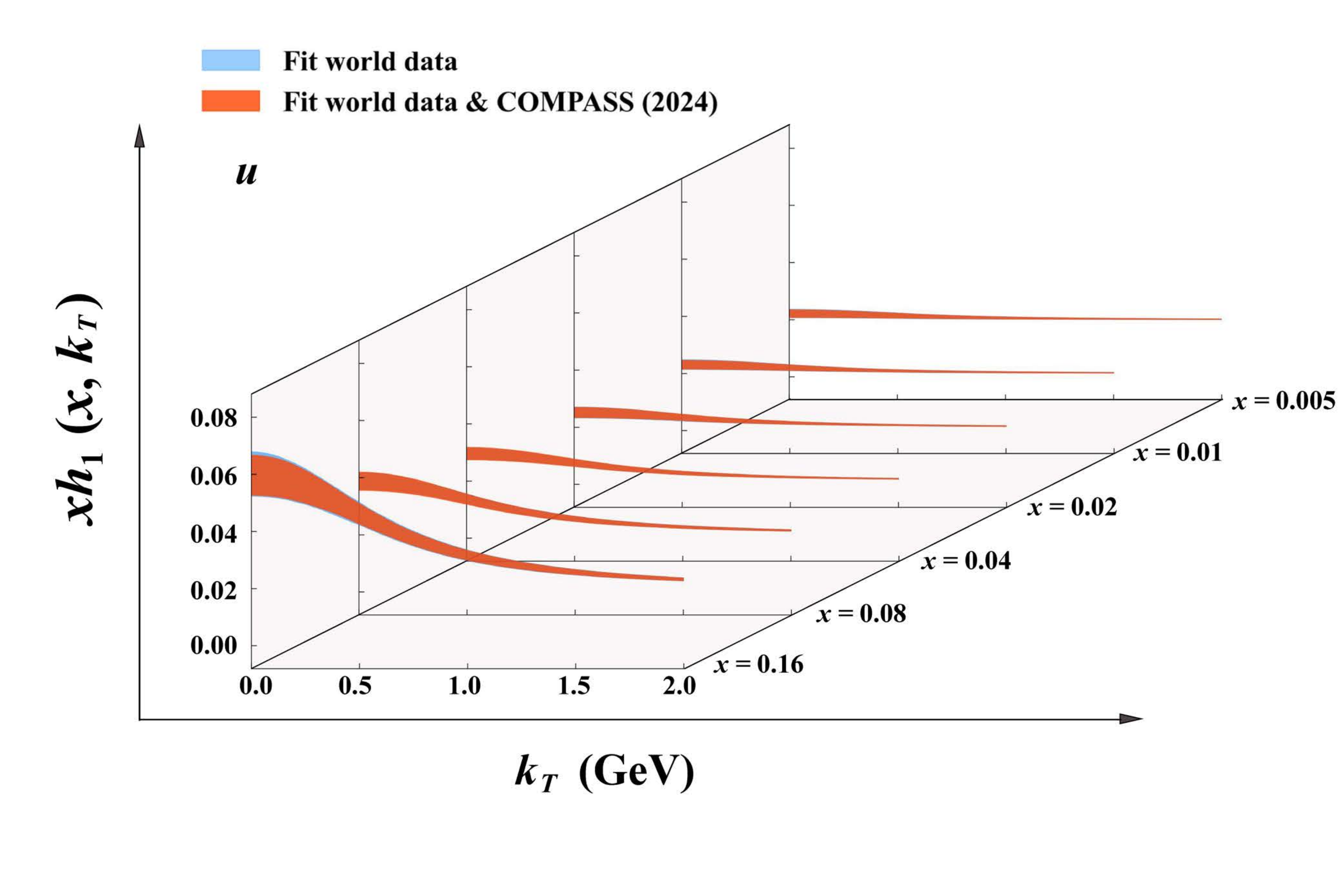}
		\includegraphics[width=0.4\textwidth]{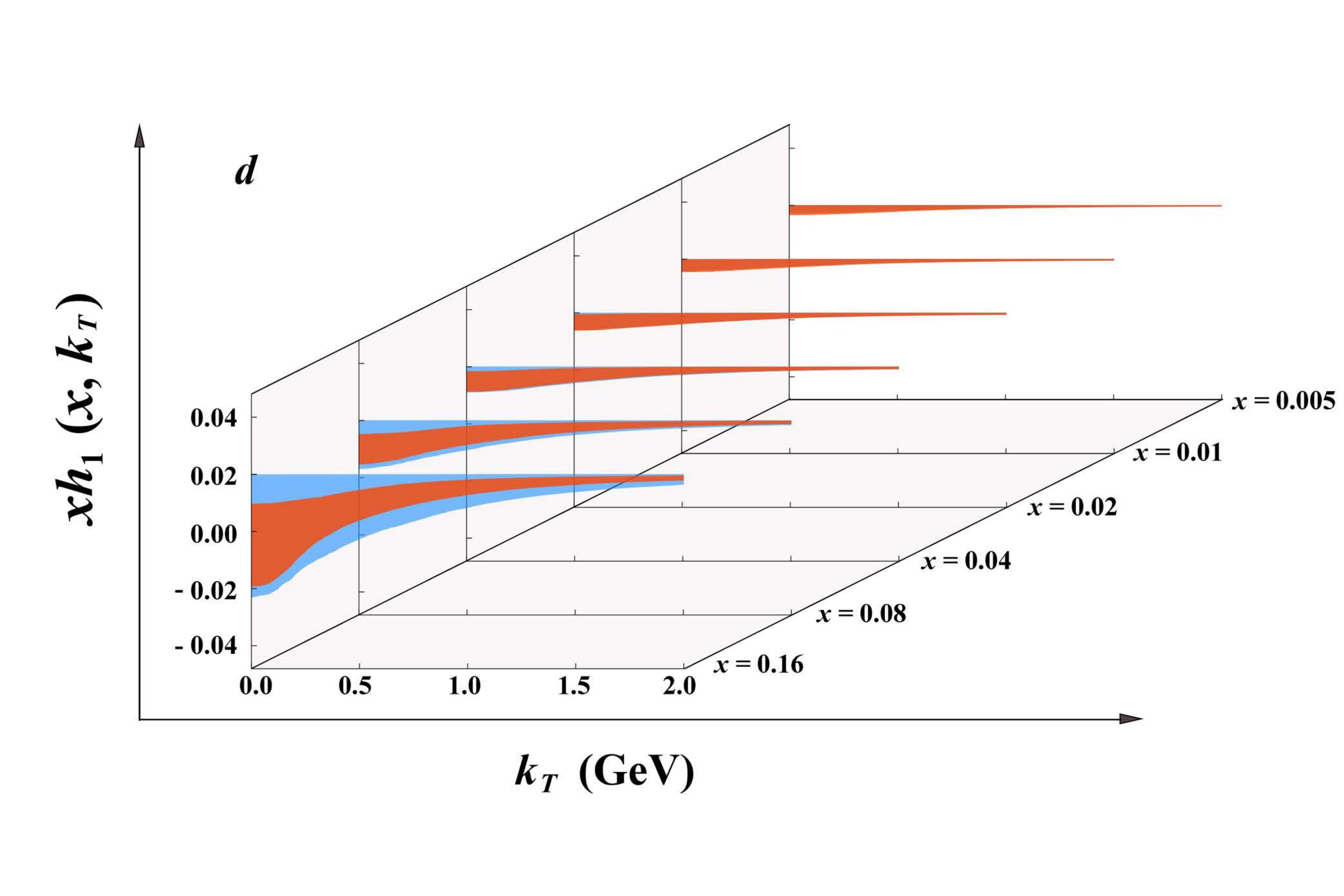}
		\includegraphics[width=0.4\textwidth]{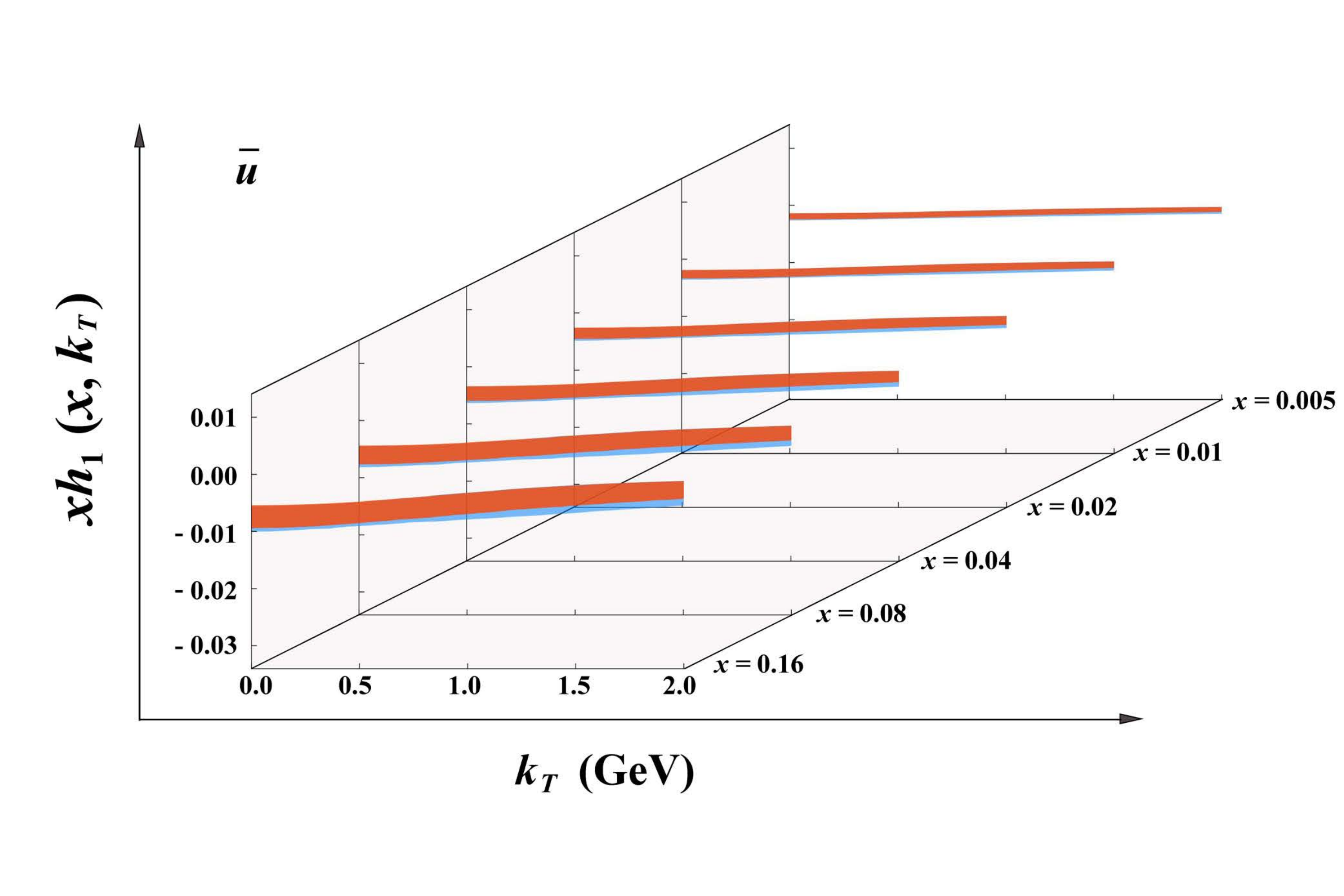}
		\includegraphics[width=0.4\textwidth]{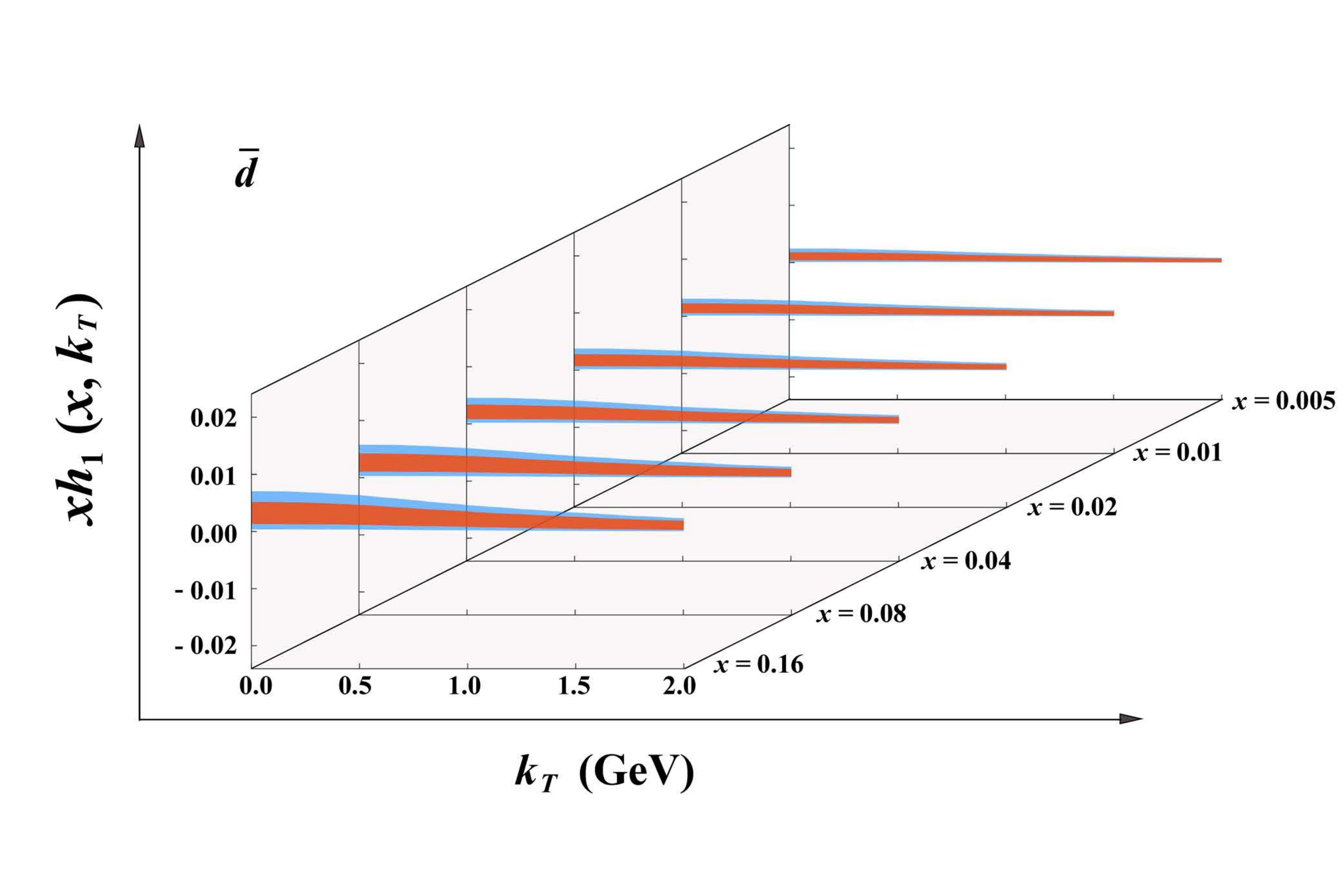}
  
		\caption{The transverse momentum distribution of the transversity functions at different $x$ values and $Q=2\, \rm GeV$. The blue (red) bands represent the uncertainties of the fit to the world data without (with) the  new COMPASS data~\cite{COMPASS:2023vhr}.}
		\label{fig:xh1x_xslices}
\end{figure*}

\begin{figure*}[htp]
		\centering
		\includegraphics[width=0.4\textwidth]{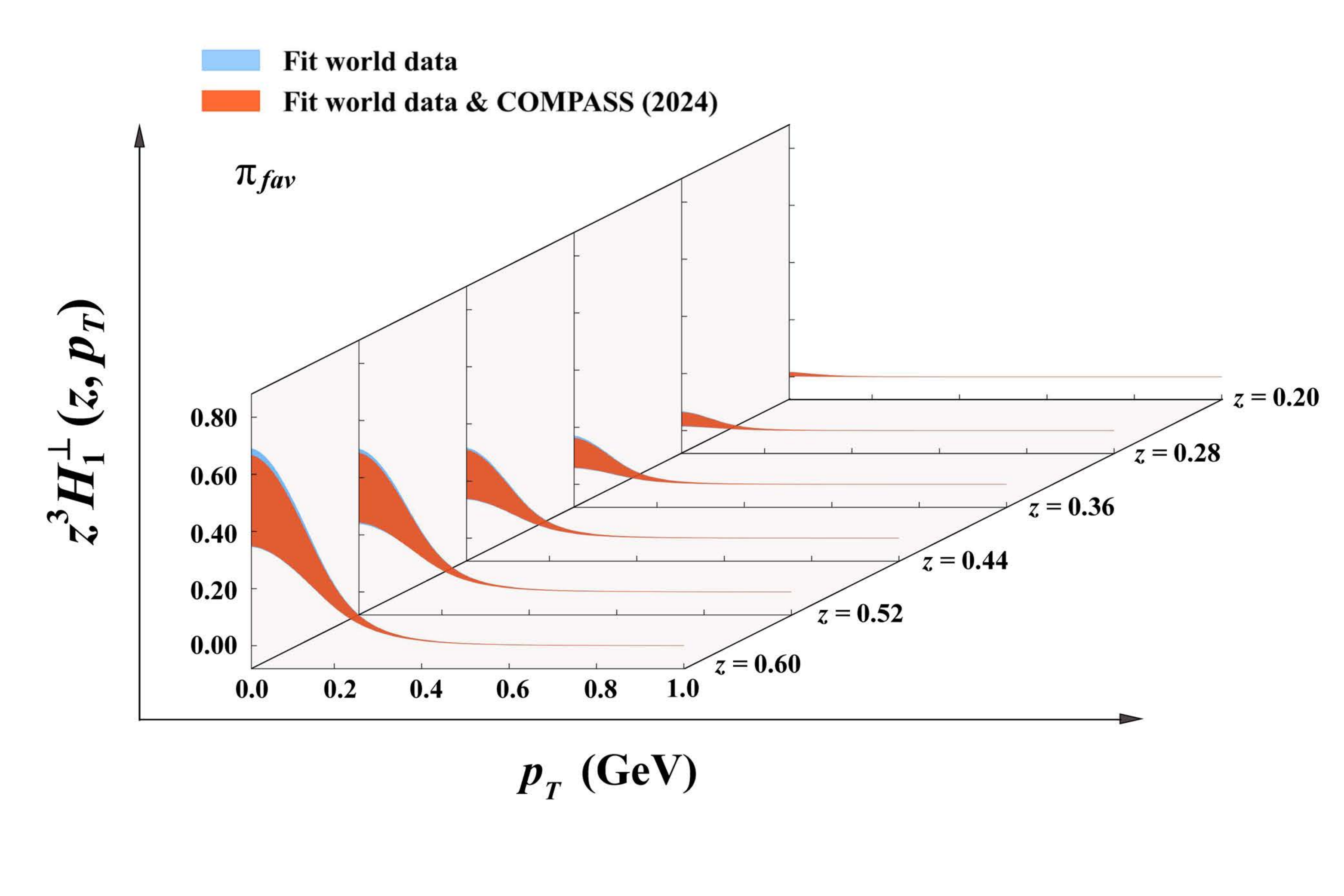}
		\includegraphics[width=0.4\textwidth]{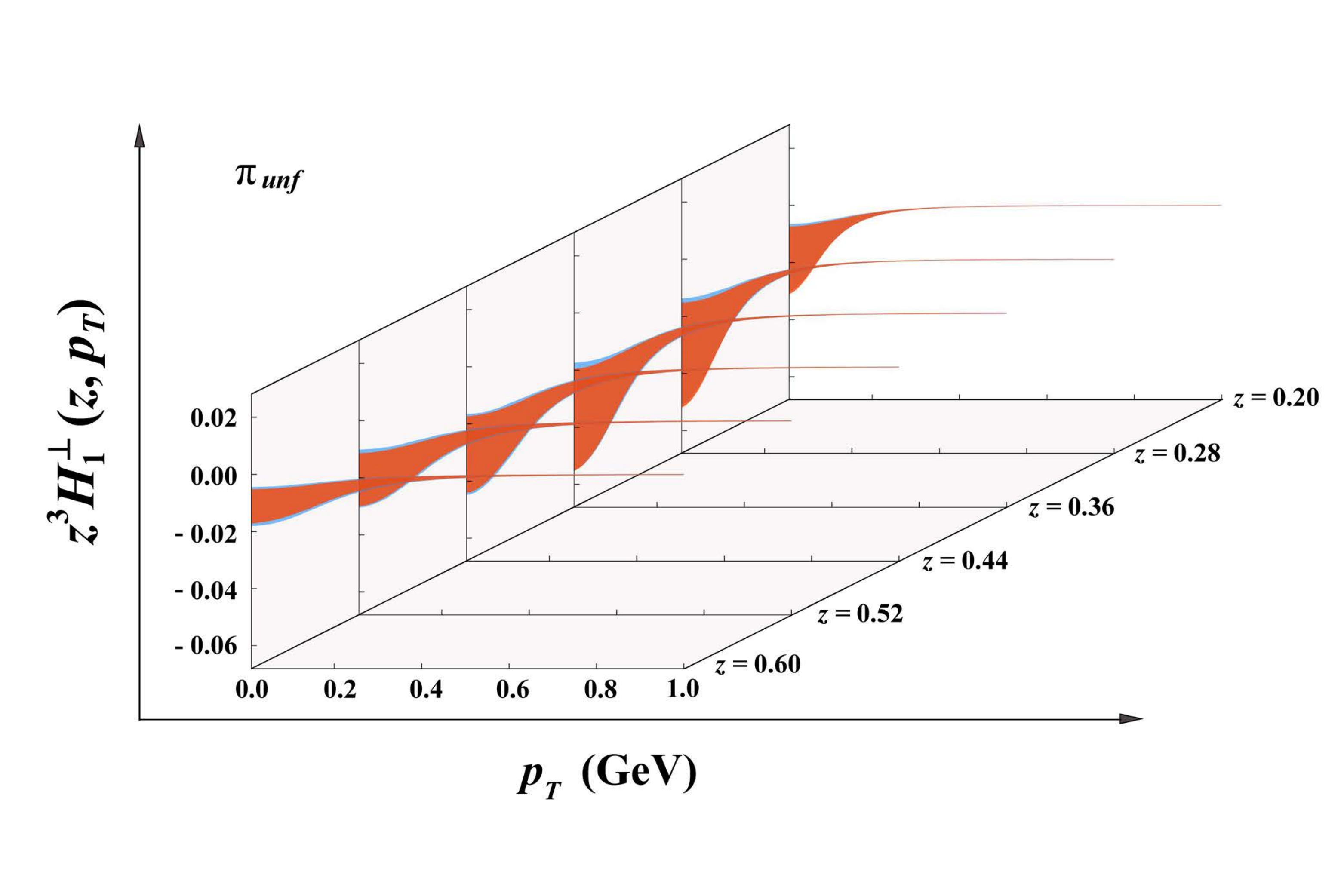}
		\includegraphics[width=0.4\textwidth]{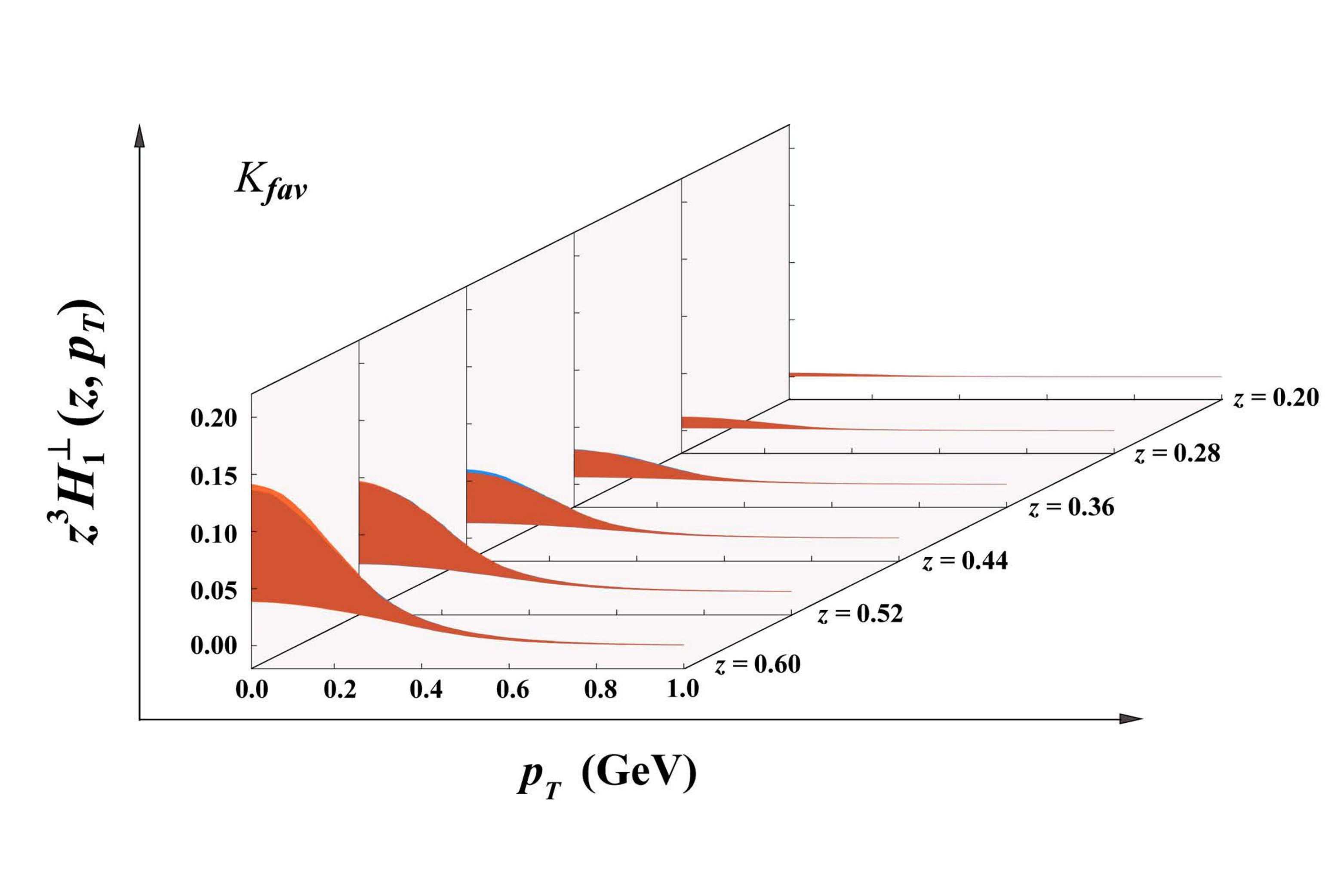}
		\includegraphics[width=0.4\textwidth]{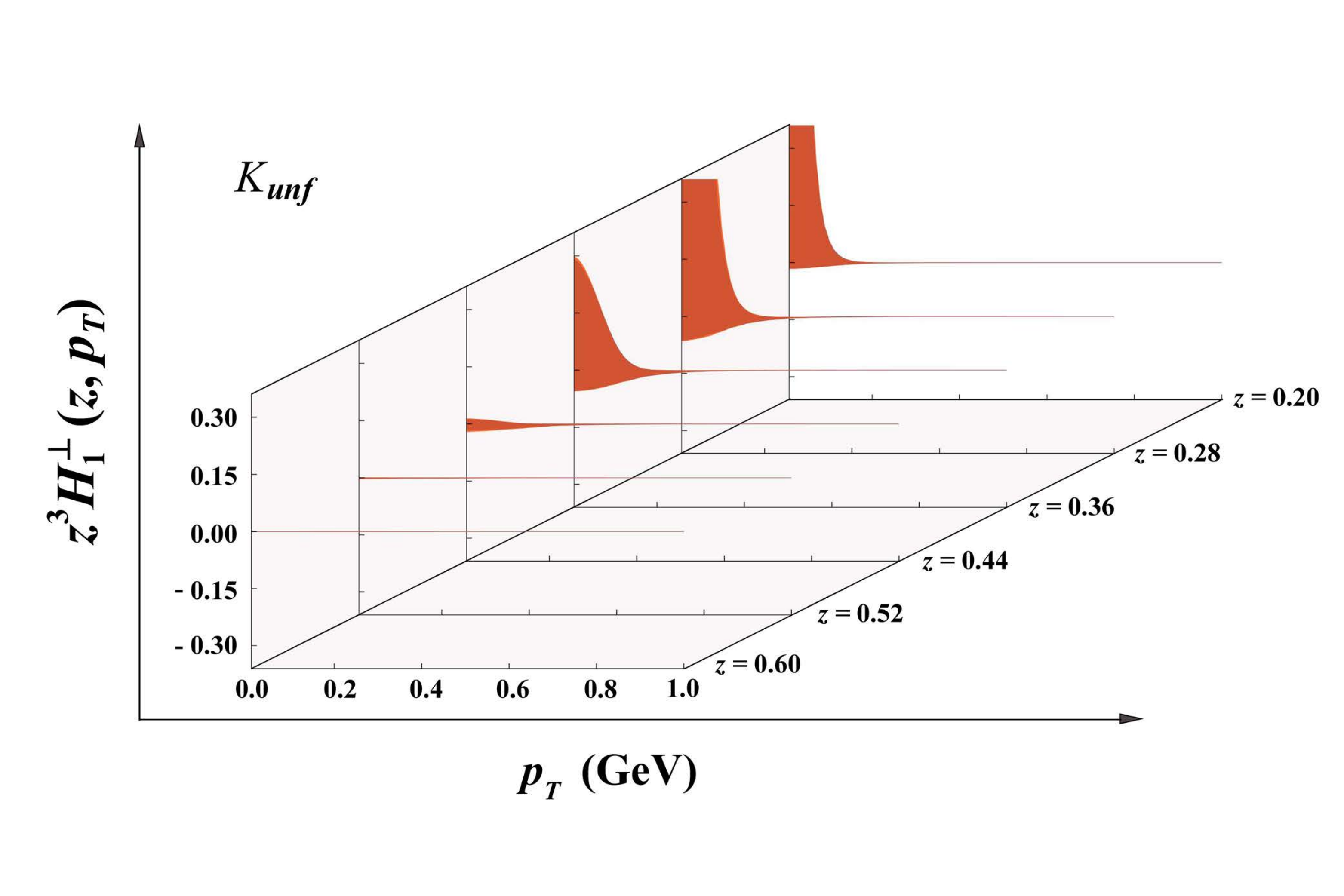}
  
		\caption{ The transverse momentum distribution of the Collins functions at different $z$ values and $Q=2\, \rm GeV$. The blue (red) bands represent the uncertainties of the fit to the world data without (with) the  new COMPASS data~\cite{COMPASS:2023vhr}.}
		\label{fig:zH1z_zslices}
\end{figure*}

The transverse momentum distribution of the Sivers, transversity, and Collins functions are shown in Figs.~\ref{fig:xfTx_xslices} to~\ref{fig:zH1z_zslices} via slices at
various $x$ or $z$ values, where $\bm{k}_{T}$ is the transverse momentum of
the quark inside the nucleon, and $\bm{p}_{T}$ is the transverse momentum of the final-state hadron with respect to the parent
quark momentum. 
In each figure, the blue bands represent the uncertainties of the fit
to the world data without the new COMPASS data~\cite{COMPASS:2023vhr}, and the red bands represent the results including the new COMPASS data. As one can observe, the incorporation of the latest COMPASS data significantly improve the precision of the Sivers and transversity functions for $d$ and $\bar{d}$ quarks. Furthermore, the integration of the most recent DY and $W^{\pm}/Z$-boson production data theoretically holds promise for improving the extraction accuracy of Sivers functions, particularly the sea quark Sivers functions. However, the limited amount of these data leads to the situation that the SIDIS process dominates the extraction of Sivers functions. Consequently, the addition of the DY and $W^{\pm}/Z$-boson production data, while valuable, does not yield a significant impact on the extracted Sivers distributions so far. Future research in this field requires the collection of high-statistics Drell-Yan (DY) data, which can be achieved by upcoming AMBER project~\cite{Quintans:2022utc} and possibly at other facilities.

\begin{figure*}[htp]
		\centering
		\includegraphics[width=0.4\textwidth]{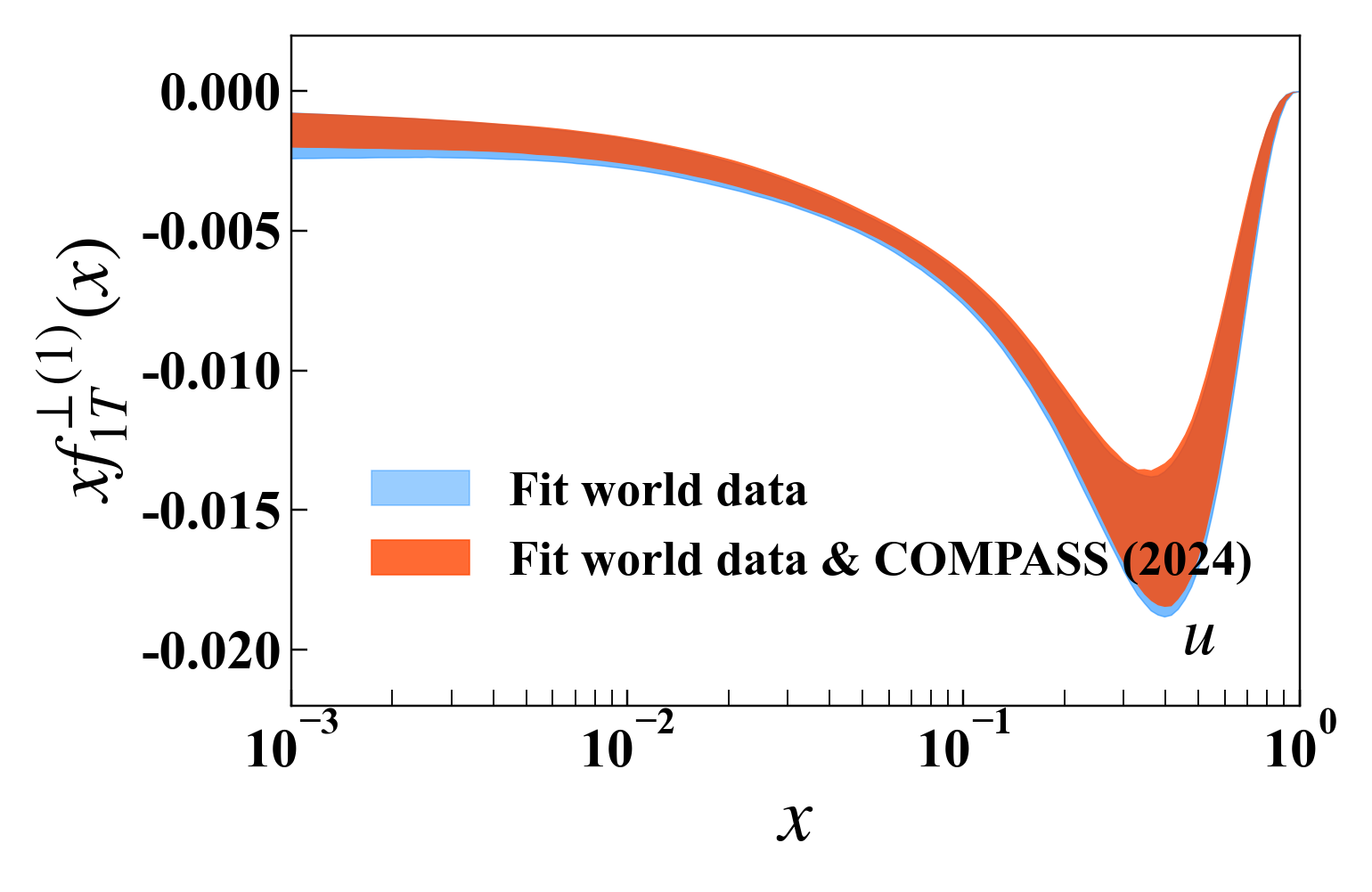}
		\includegraphics[width=0.4\textwidth]{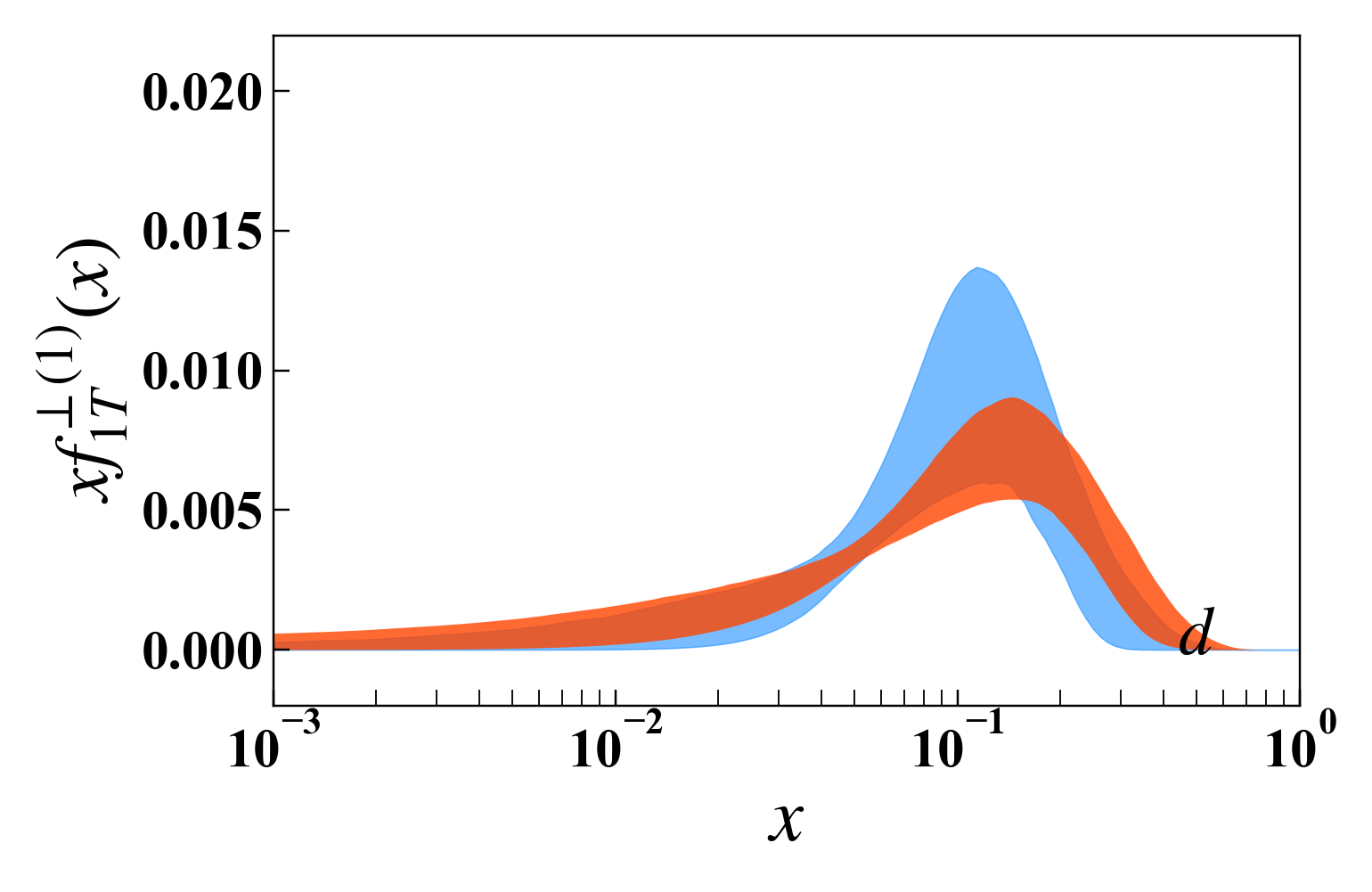}
		\includegraphics[width=0.4\textwidth]{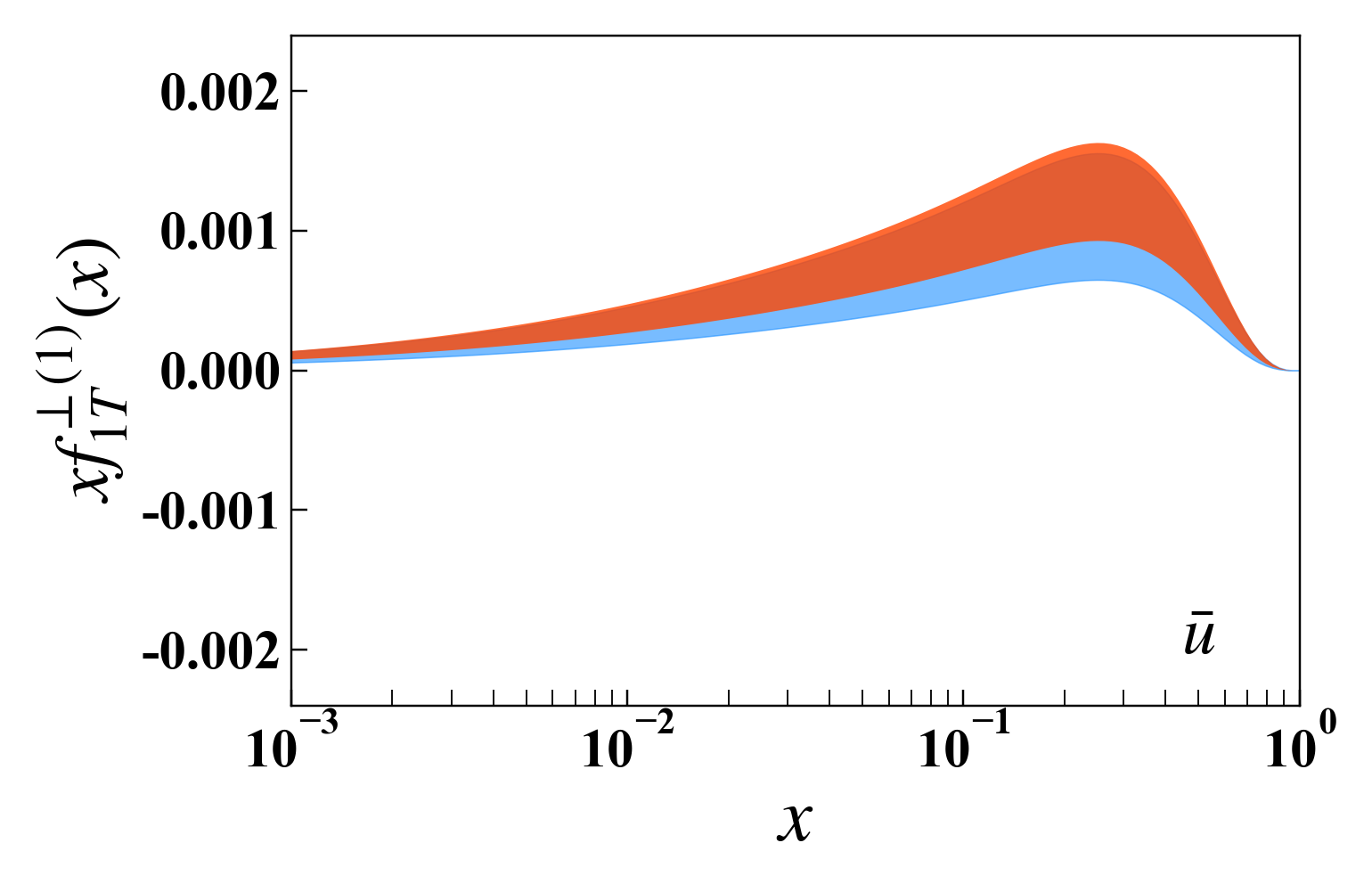}
		\includegraphics[width=0.4\textwidth]{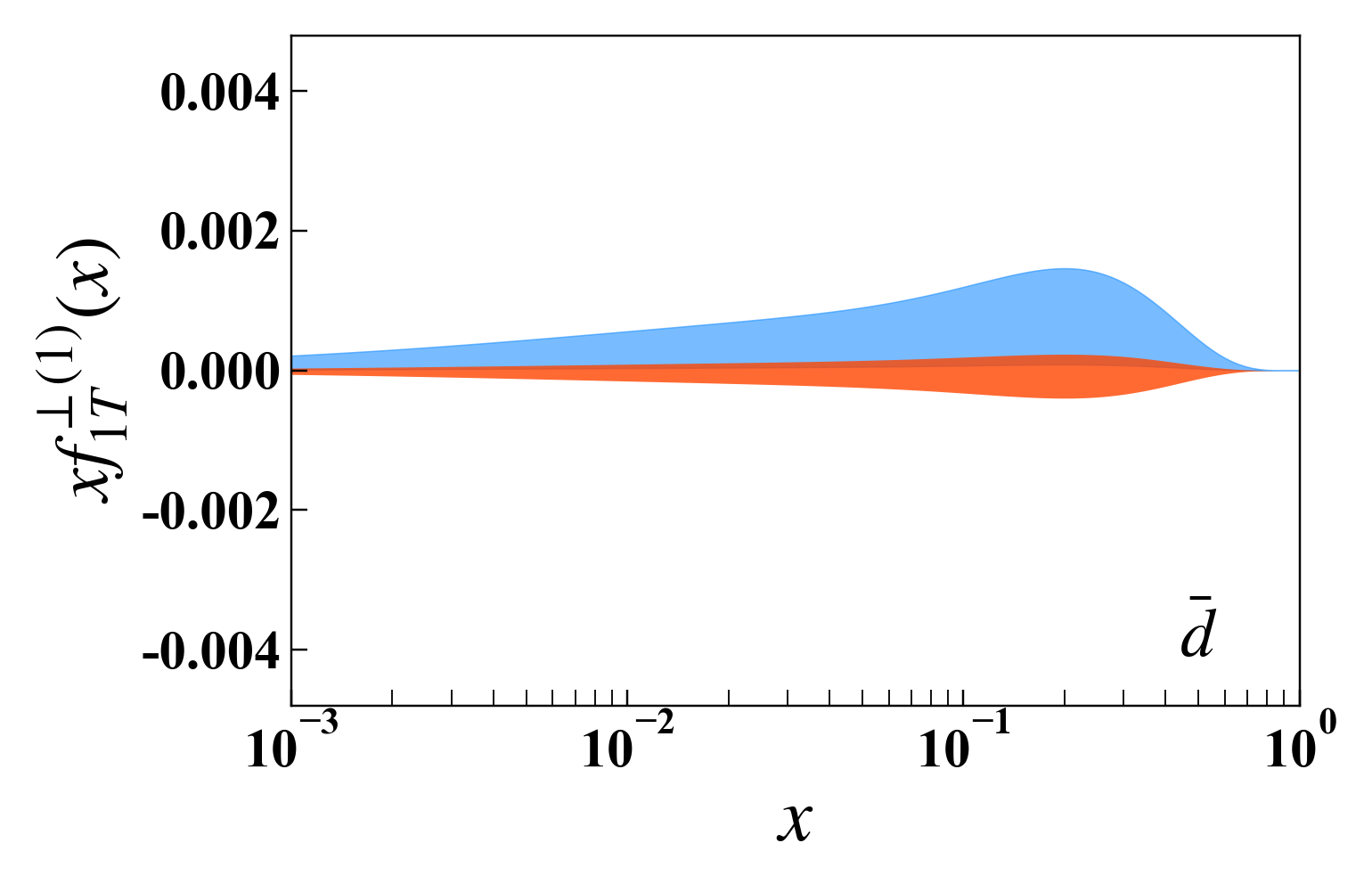}
		\caption{The first transverse moment of Sivers functions as defined in Eq.~(\ref{eq:xfTx}) with $k_T^{\text{cut}} = Q \times \delta_{\text{cut}}$ ($Q=2\, \rm GeV, \delta_{\text{cut}}=1 $). The blue (red) bands represent the uncertainties of the fit to the world data without (with) the  new COMPASS data~\cite{COMPASS:2023vhr}.}
		\label{fig:xfTx}
\end{figure*}

\begin{figure*}[htp]
		\centering
		\includegraphics[width=0.4\textwidth]{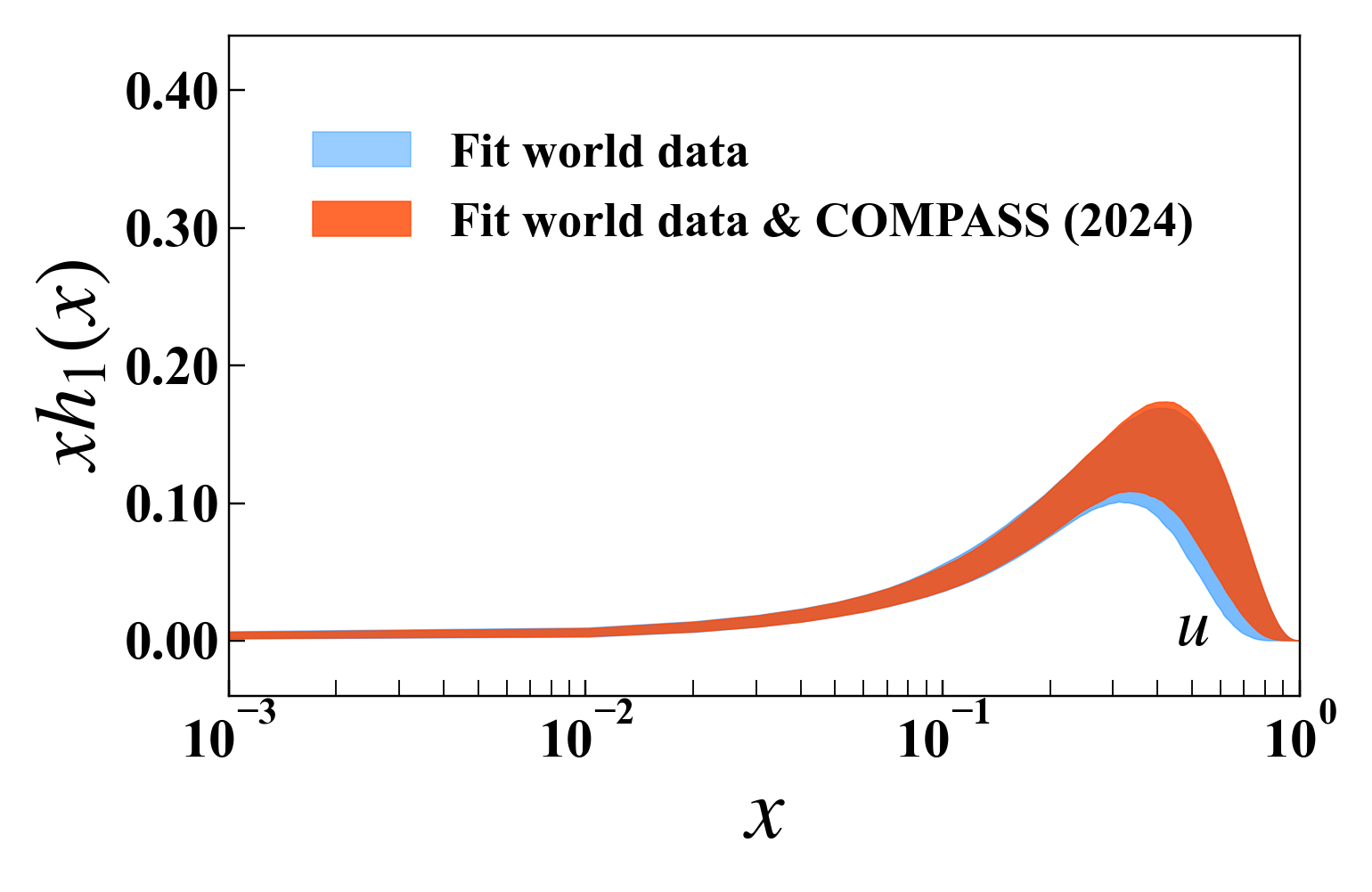}
		\includegraphics[width=0.4\textwidth]{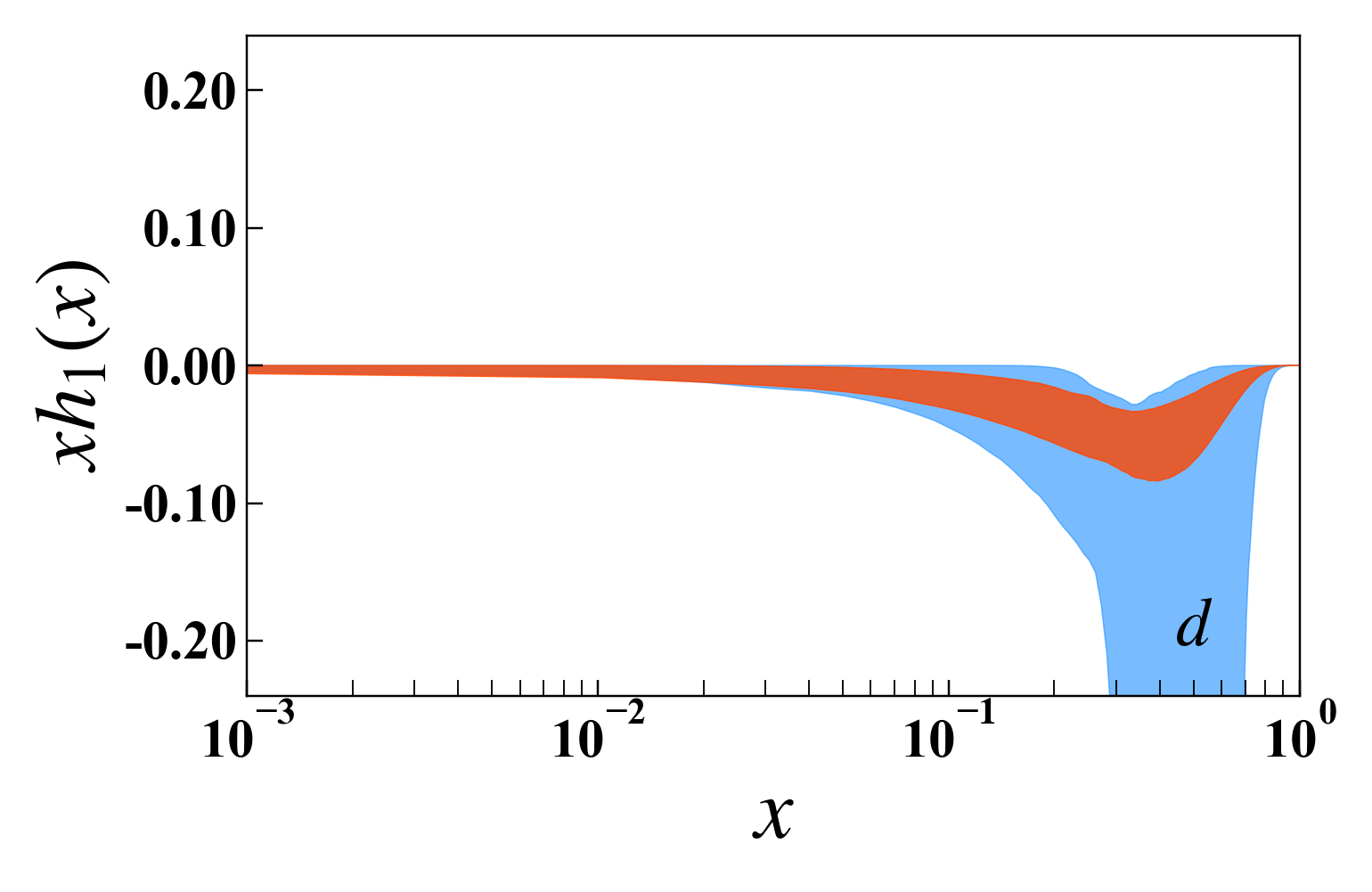}
		\includegraphics[width=0.4\textwidth]{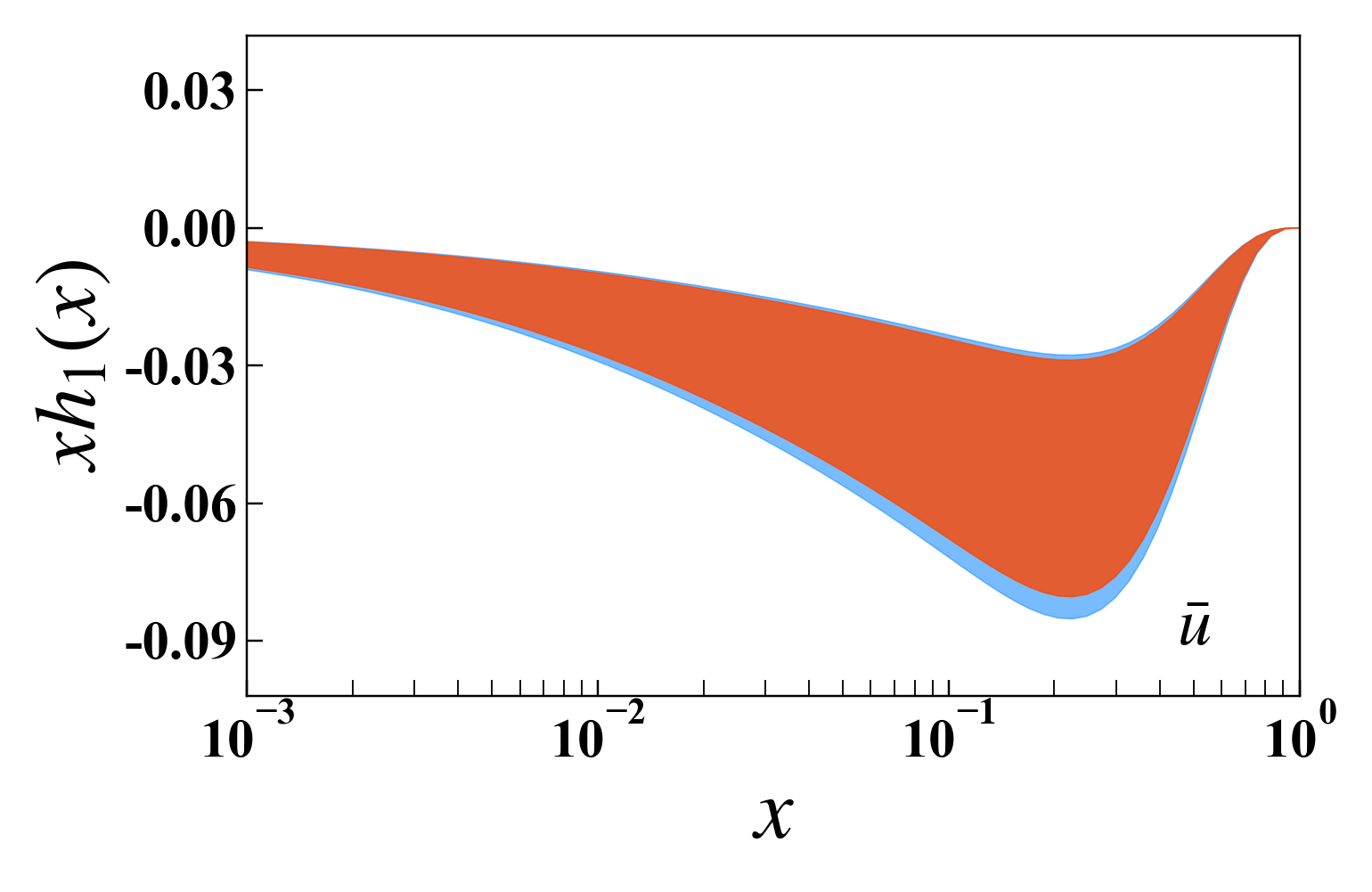}
		\includegraphics[width=0.4\textwidth]{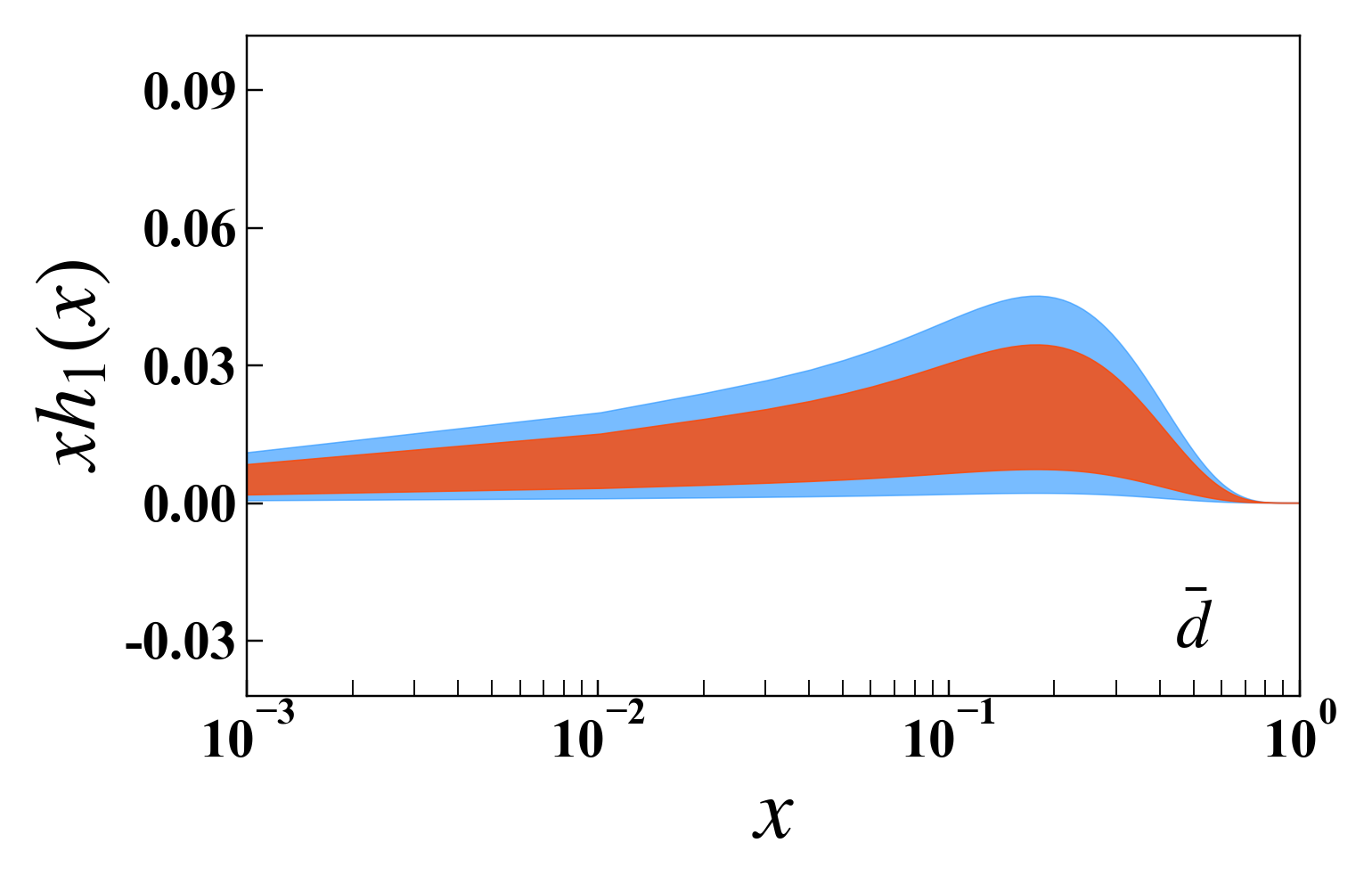} 
		\caption{Transverse momentum integrated transversity functions as defined in Eq.~(\ref{eq:xh1x}) with $ k_T^{\text{cut}} = Q  \times \delta_{\text{cut}}$ ($Q=2\, \rm GeV, \delta_{\text{cut}}=1$). The blue (red) bands represent the uncertainties of the fit to the world data without (with) the  new COMPASS data~\cite{COMPASS:2023vhr}.  
       }
		\label{fig:xh1x}
\end{figure*}

\begin{figure*}[htp]
		\centering
		\includegraphics[width=0.4\textwidth]{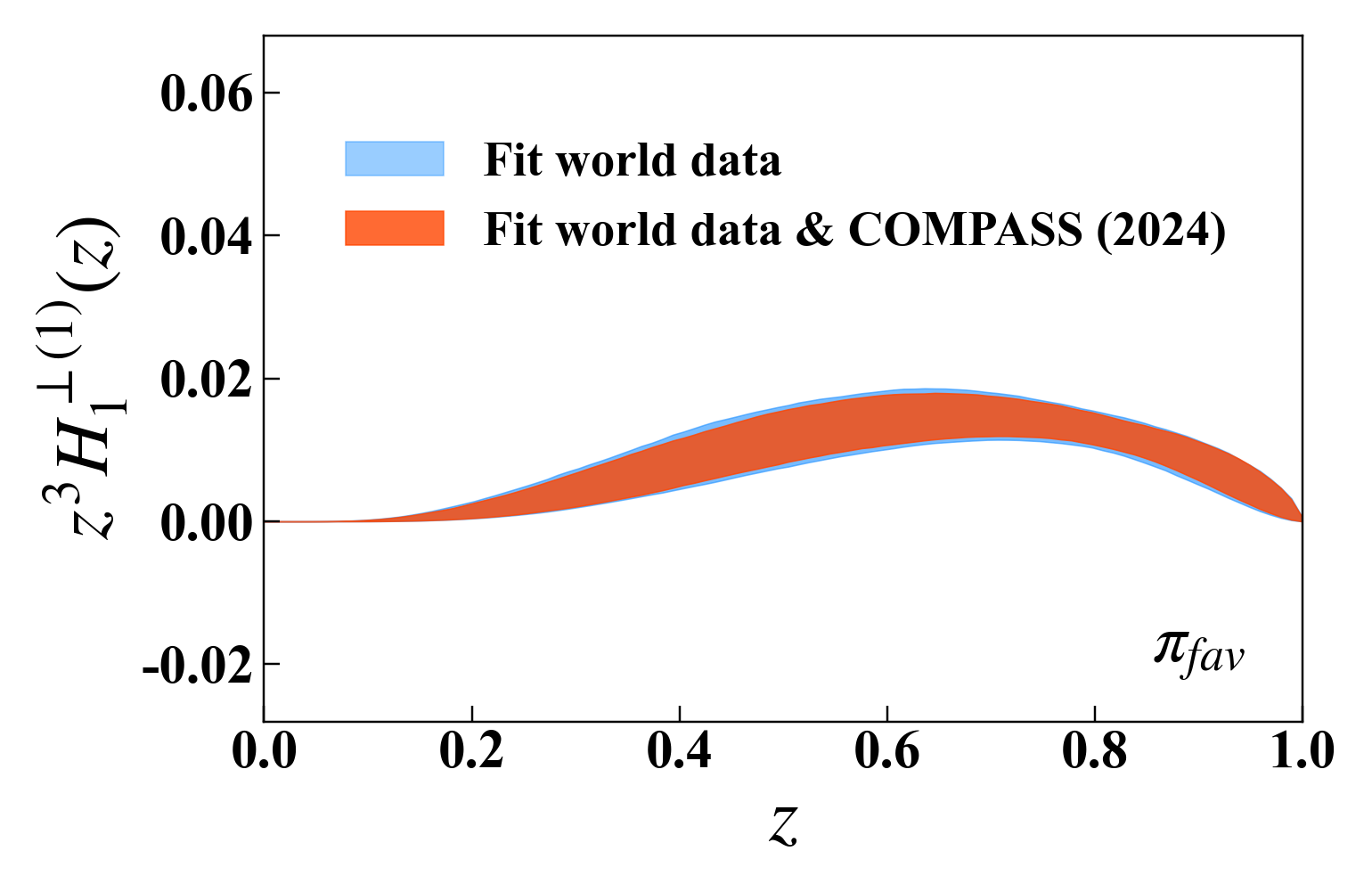}
		\includegraphics[width=0.4\textwidth]{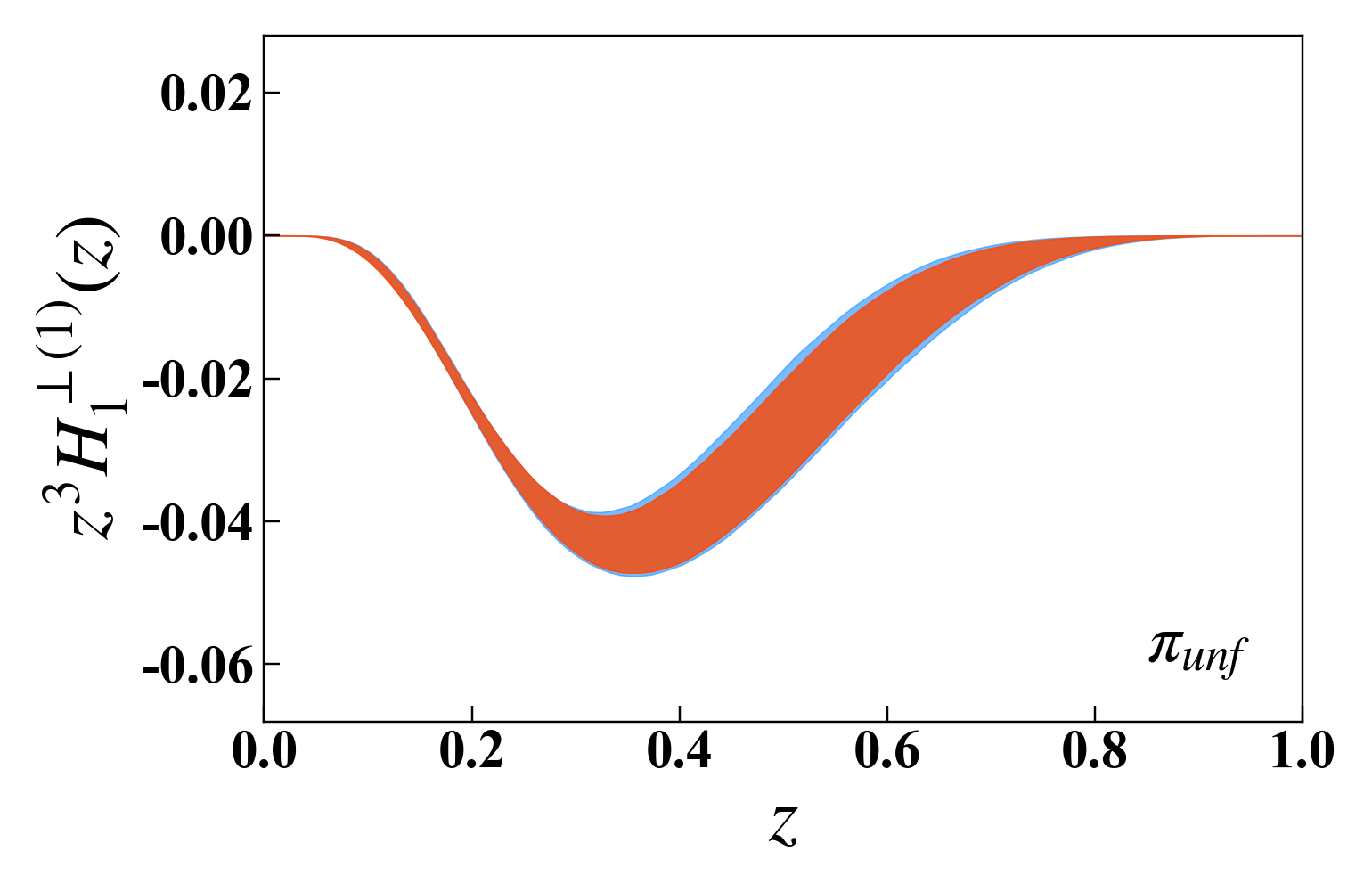}
		\includegraphics[width=0.4\textwidth]{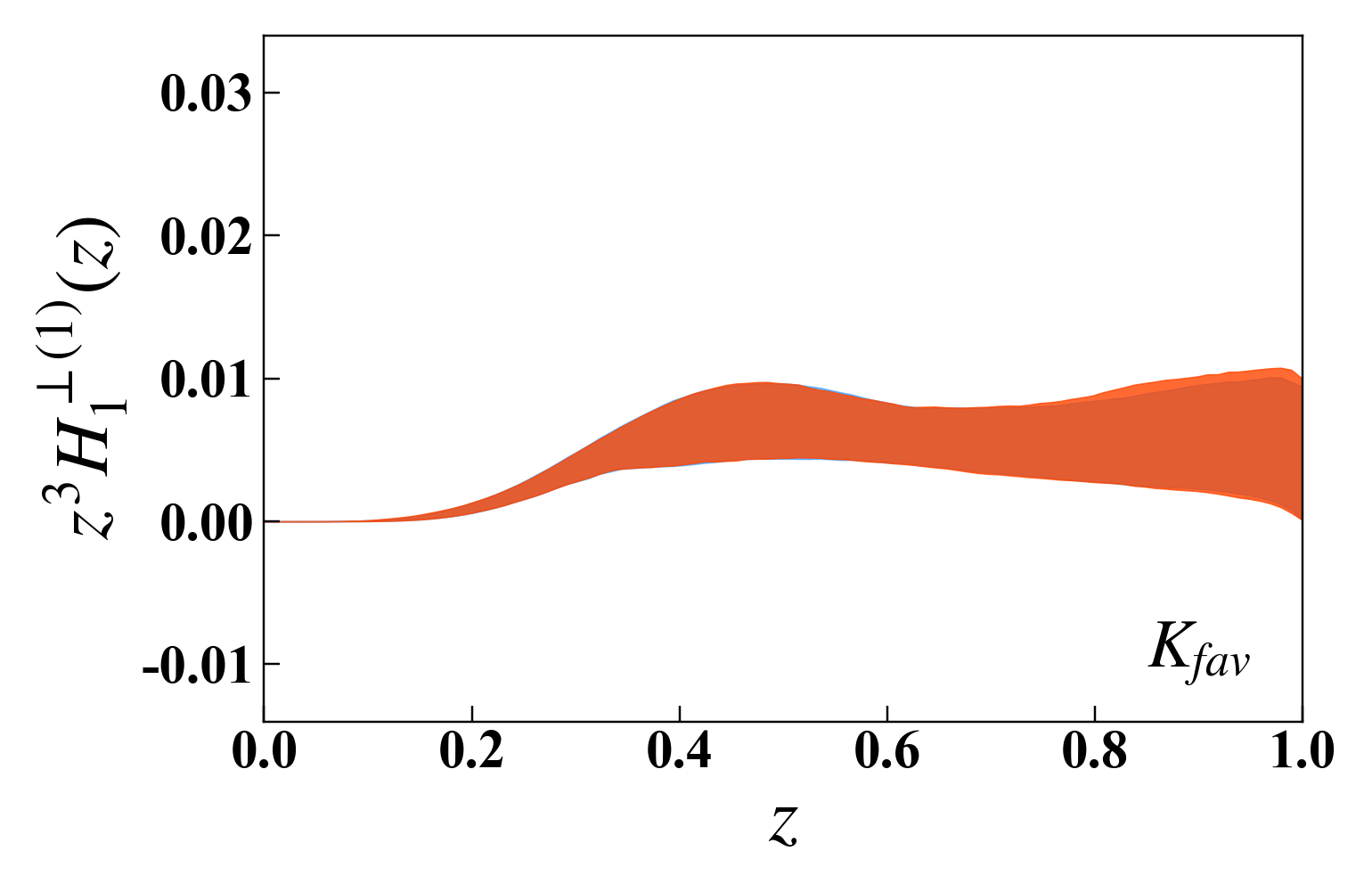}
		\includegraphics[width=0.4\textwidth]{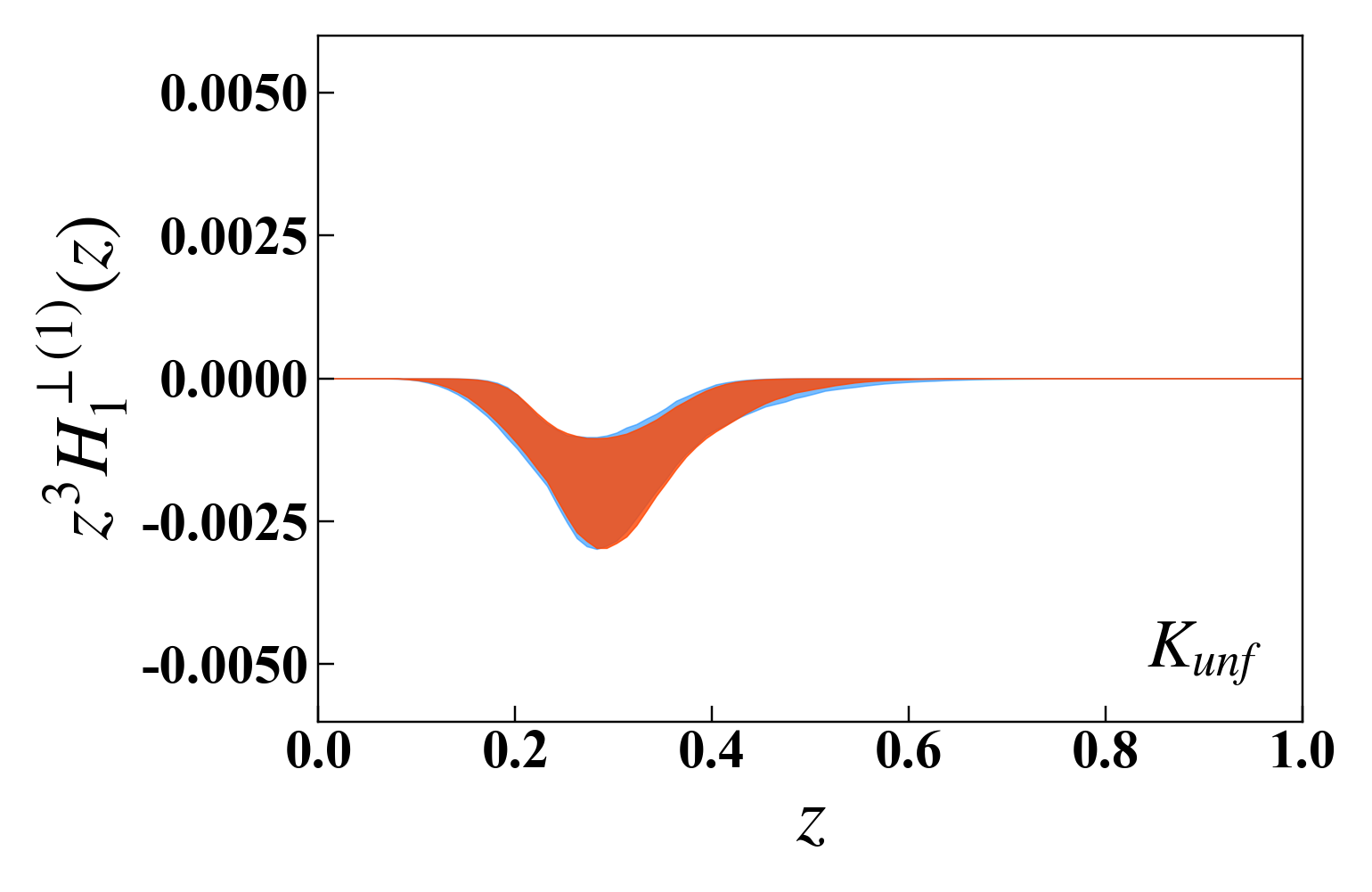}  
		\caption{The first transverse moment of Collins fragmentation functions as defined in Eq.~(\ref{eq:zH1z}) with  $ p_T^{\text{cut}} = Q \times \delta_{\text{cut}} / z $ ($Q=2\, \rm GeV, \delta_{\text{cut}}=1$). The blue (red) bands represent the uncertainties of the fit to the world data without (with) the  new COMPASS data~\cite{COMPASS:2023vhr}.}
		\label{fig:zH1z}
\end{figure*}

\begin{figure*}[htp]
		\centering
  		\includegraphics[width=0.45\textwidth]{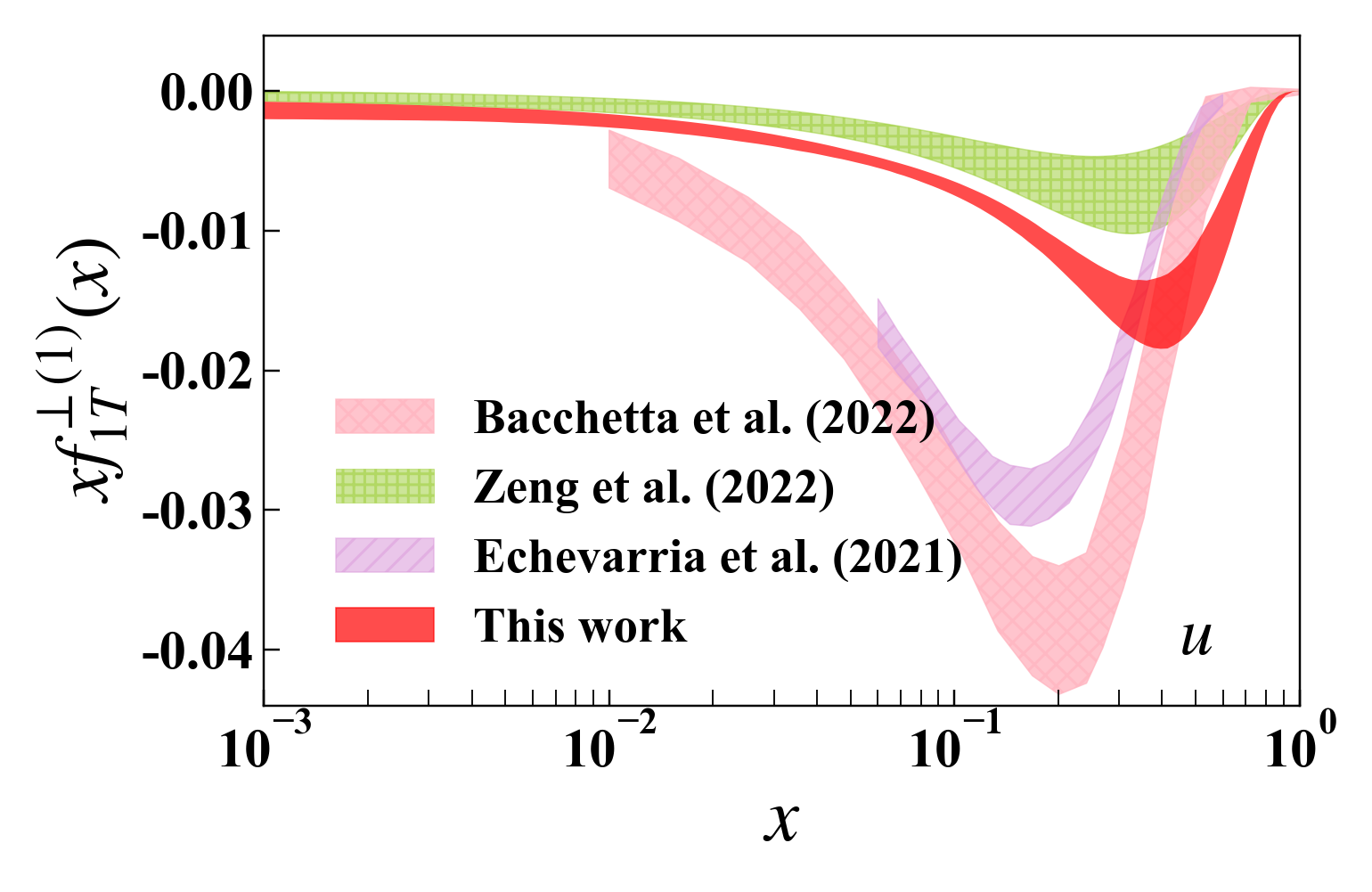}
		\includegraphics[width=0.45\textwidth]{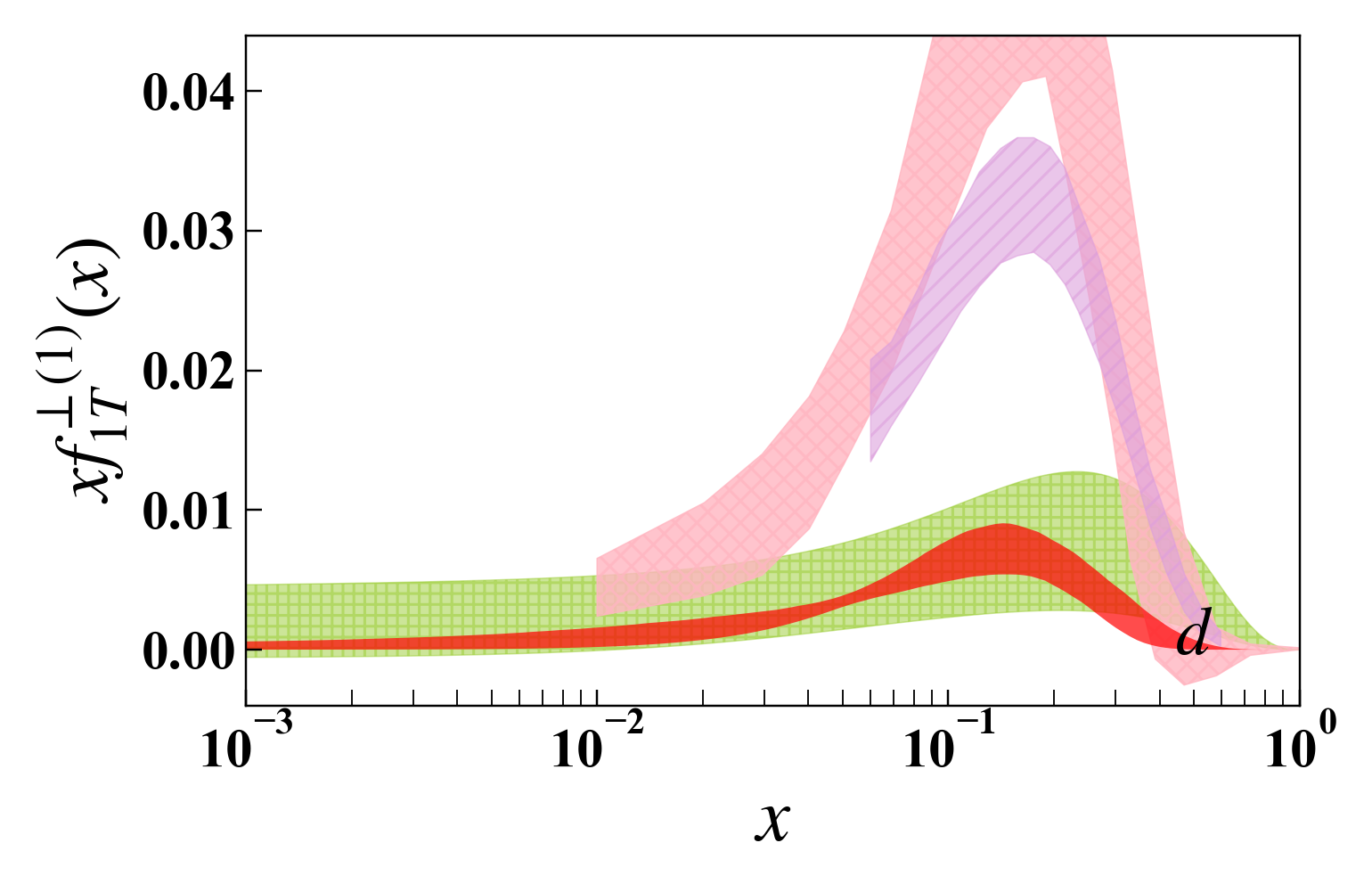}
		\caption{ The extracted first transverse moment of the Sivers functions ($Q=2\, \rm GeV$).  Our
        results are compared with Zeng et al.~\cite{Zeng:2022lbo} ($Q=2\, \rm GeV$), Echevarria et al.~\cite{Echevarria:2020hpy} ($Q=\sqrt{1.9} \, \rm GeV$), and Bacchetta et al.~\cite{ Bacchetta:2020gko} ($Q=2\, \rm GeV$). In this work, the transverse momentum integral as defined Eq.~(\ref{eq:xfTx}) is truncated at $k_T^{\text{cut}} = Q \times \delta_{\text{cut}}$ with $\delta_{\rm cut}=1$. In Zeng et al.~\cite{Zeng:2022lbo}, the transverse momentum integral is truncated with $\delta_{\rm cut}=0.3$.}
        \label{fig:sivers_compare}
\end{figure*}

\begin{figure*}[htp]
		\centering
            \includegraphics[width=0.45\textwidth]{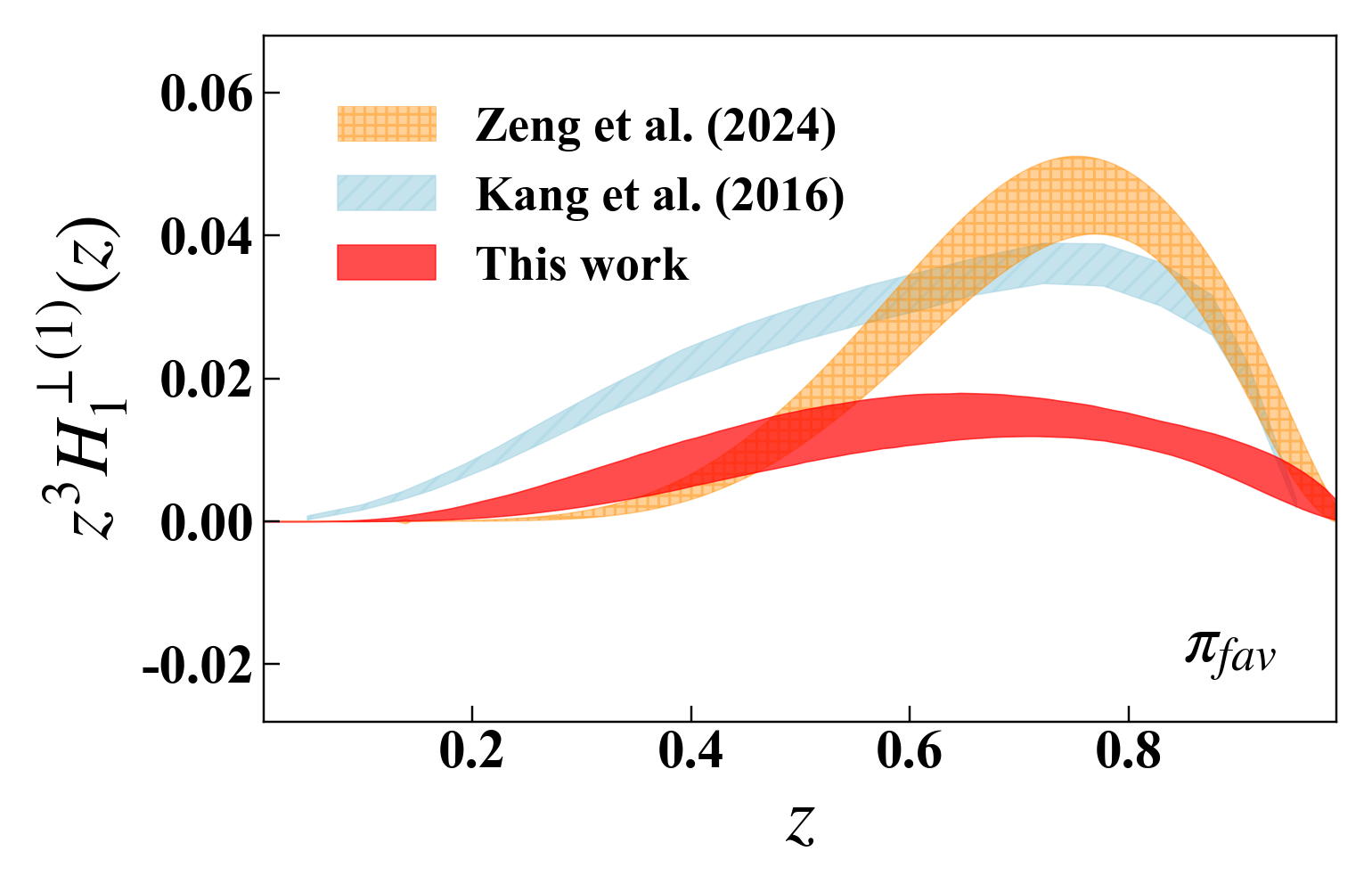}
		\includegraphics[width=0.45\textwidth]{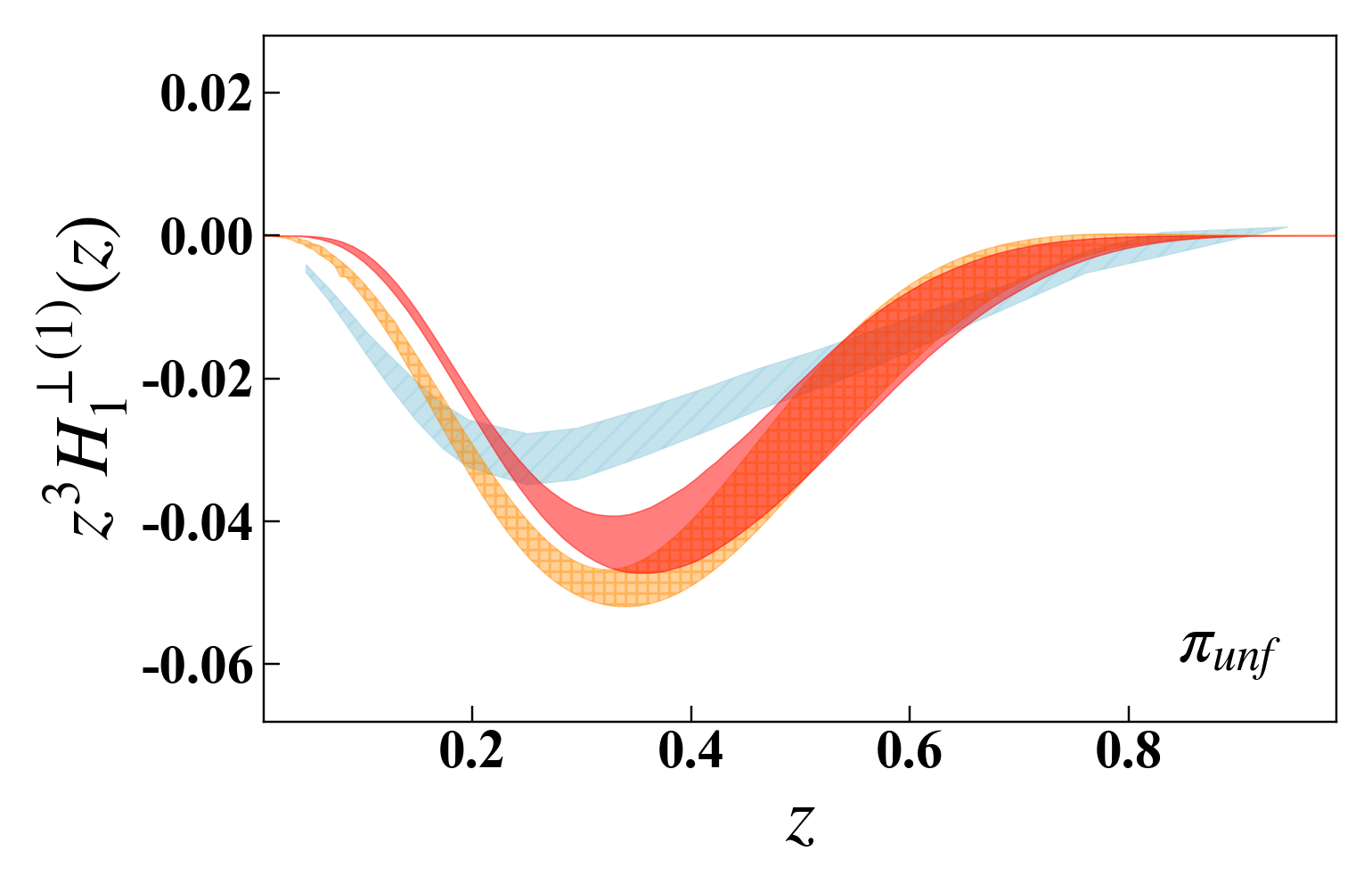}
		\caption{The extracted first transverse moment of the Collins fragmentation function ($Q=2\, \rm GeV$). Our
        results are compared with with Zeng et al.~\cite{Zeng:2023nnb} ($Q=2\, \rm GeV$) and Kang et al.~\cite{Kang:2015msa} ($Q=\sqrt{10}\, \rm GeV$). In this work, the transverse momentum integral as defined Eq.~(\ref{eq:zH1z}) is truncated at $p_T^{\text{cut}}/z = Q \times \delta_{\text{cut}}$ with $\delta_{\rm cut}=1$. In Zeng et al.~\cite{Zeng:2023nnb}, the integral is truncated at $p_T^{\rm cut} = 1 \,\rm GeV$.}
        \label{fig:collins_compare}
\end{figure*}

The first transverse moments of the Sivers function $f_{1T}^{\perp(1)}(x)$, the Collins FF $H_1^{\perp(1)}(z)$ and the transversity function $h_1(x)$ are respectively defined as
\begin{align}
    f_{1T}^{\perp (1)}(x)
    &=  \int_0^{k_{T}^{\text{cut}}} d^2 {\bm k}_{T}\, \frac{{\bm k}_{T}^2}{2M^2}f_{1T}^{\perp}(x, { k}_{T}),\label{eq:xfTx}\\
    {H}_1^{\perp(1)}(z)&=\int_0^{p_{T}^{\text{cut}}} d^2\bm{p}_{T} \frac{p_{T}^2}{2z^2M_h^2}H_{1}^{\perp}(z, p_T) \label{eq:zH1z} ,\\  
    h_{1}(x)&=\int_0^{k_{T}^{\text{cut}}} d^2\bm{k}_{T} h_{1}(x, k_{T}) .\label{eq:xh1x}
\end{align}
We note that the TMD formalism is not valid at large transverse momentum $k_T$ (or $p_T/z$) region, and hence we apply a $k_T^{\text{cut}}=Q\times \delta_{\text{cut}}$ ($p_T^{\text{cut}}/z=Q\times \delta_{\text{cut}}$) for the integration. The fitting results of  $f_{1T}^{\perp(1)}(x)$, $h_1(x)$ and $H_1^{\perp(1)}(z)$ are respectively shown in Figs.~\ref{fig:xfTx} to~\ref{fig:zH1z} at $Q = 2\,\rm GeV$ with $\delta_{\rm cut}=1$. The comparisons with previously results are shown in Figs.~\ref{fig:sivers_compare} to~\ref{fig:collins_compare}. 
Note that the studies are slighted different. The $f_{1T}^{\perp(1)}(x)$, $h_1(x)$ and $H_1^{\perp(1)}(z)$ are directly extracted from the fit of data as a collinear function in~\cite{Echevarria:2020hpy} by Echevarria et al. and~\cite{Kang:2015msa} by Kang et al., which do not include the cut $k_T^{\text{cut}} (p_T^{\text{cut}}$), while in~\cite{ Bacchetta:2020gko} by Bacchetta et al., the first transverse moment of the Sivers function $f_{1T}^{\perp(1)}(x)$ is obtained with $b_{\text{min}}$, which effectively introduced a $k_T^{\text{cut}}$ indirectly.

\begin{figure}[htp]
    \centering
    \includegraphics[width=1.0\columnwidth]{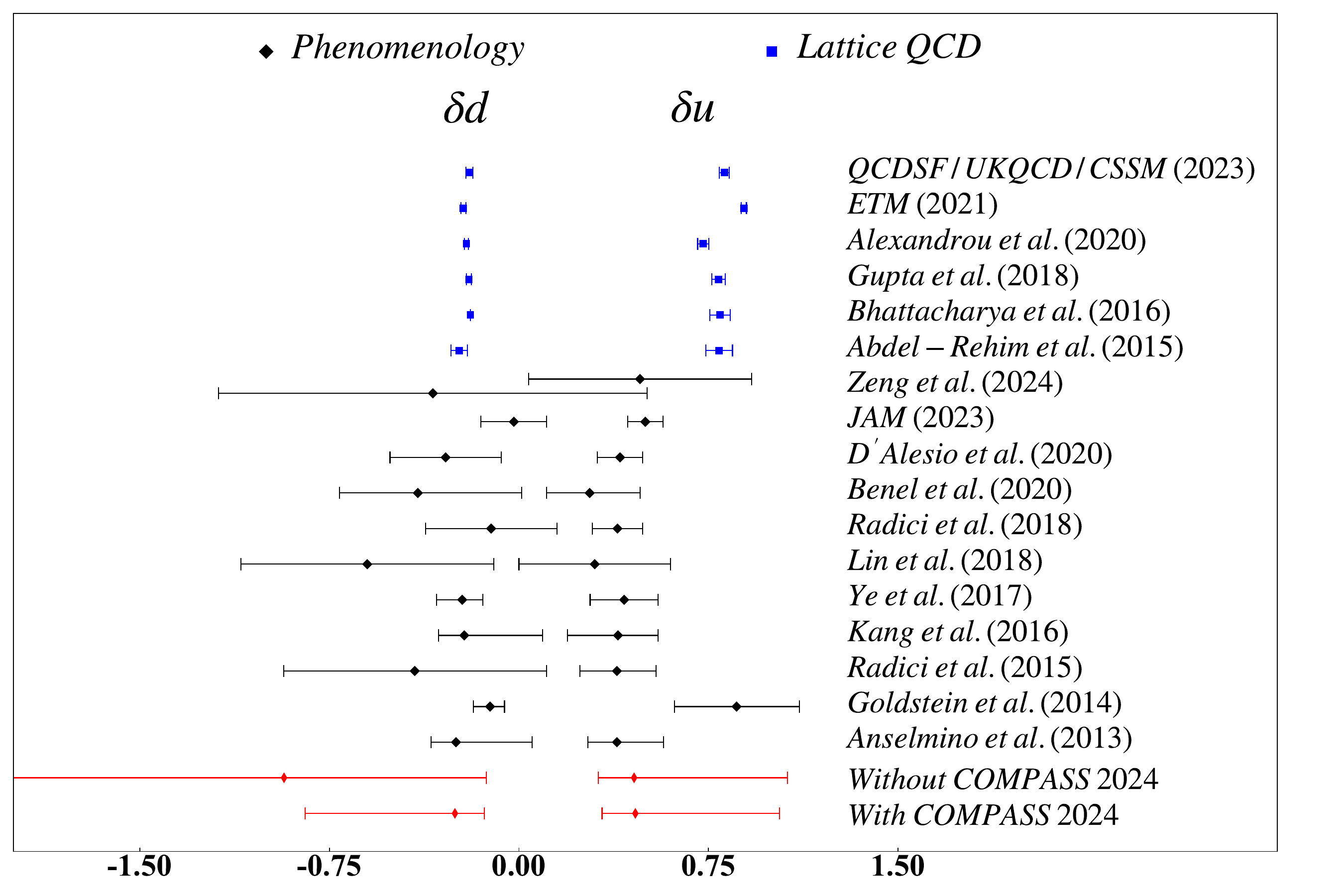}
    \caption{Tensor charge 
for $\delta u$ and $\delta d$ from this analysis with the error bars representing 68\% C.L.,
along with the results from lattice QCD calculations~\cite{QCDSFUKQCDCSSM:2023qlx,Alexandrou:2021oih,Alexandrou:2019brg,Gupta:2018qil,Bhattacharya:2016zcn,Abdel-Rehim:2015owa}, and phenomenological extractions~\cite{Zeng:2023nnb,Cocuzza:2023vqs,Ye:2016prn, Kang:2015msa,Radici:2015mwa,Goldstein:2014aja,Anselmino:2013vqa,Lin:2017stx,Radici:2018iag,Benel:2019mcq,DAlesio:2020vtw}.}
    \label{fig:gugd}
\end{figure}

The tensor charge can be evaluated from the integral of the transversity distributions as
\begin{align}
    \delta u&=\int_0^1 dx (h_1^u(x)-h_1^{\bar{u}}(x)),\\
    \delta d&=\int_0^1 dx (h_1^d(x)-h_1^{\bar{d}}(x)),
\end{align}
and the isovector combination is given by
\begin{align}
    \textsl{g}_T=\delta u-\delta d.
\end{align}
The extracted tensor charges from our analysis with (without) COMPASS new data~\cite{ COMPASS:2023vhr} are $\textsl{g}_T = 0.80_{-0.31}^{+1.69} (1.73_{-1.23}^{+6.38})$. The comparison of $\delta u$ and $\delta d$ with the results from previous phenomenological studies and lattice calculations as shown in~Fig.~\ref{fig:gugd}. 
Based on these results, it is clear that the latest COMPASS data significantly improve the precision of the determination of transversity and Sivers distributions of $d$ and $\bar d$ quarks, and the tensor charge $\delta d$. The new results in this analysis also support the finding that the sea quark can have nonzero transversity distribution~\cite{Zeng:2023nnb}.

\begin{figure*}[htp]
	\centering
    \includegraphics[width=0.45\textwidth]{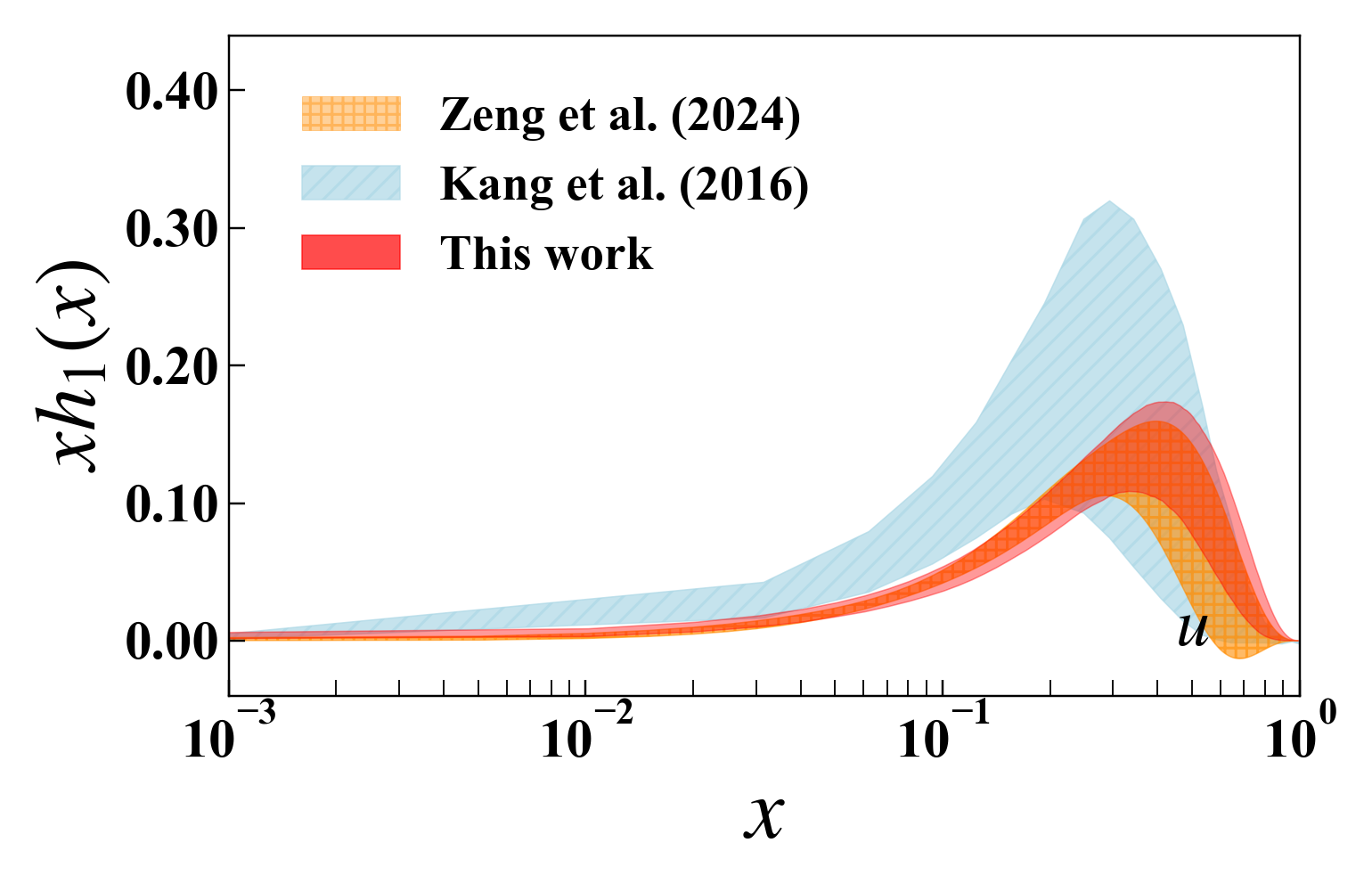}
	\includegraphics[width=0.45\textwidth]{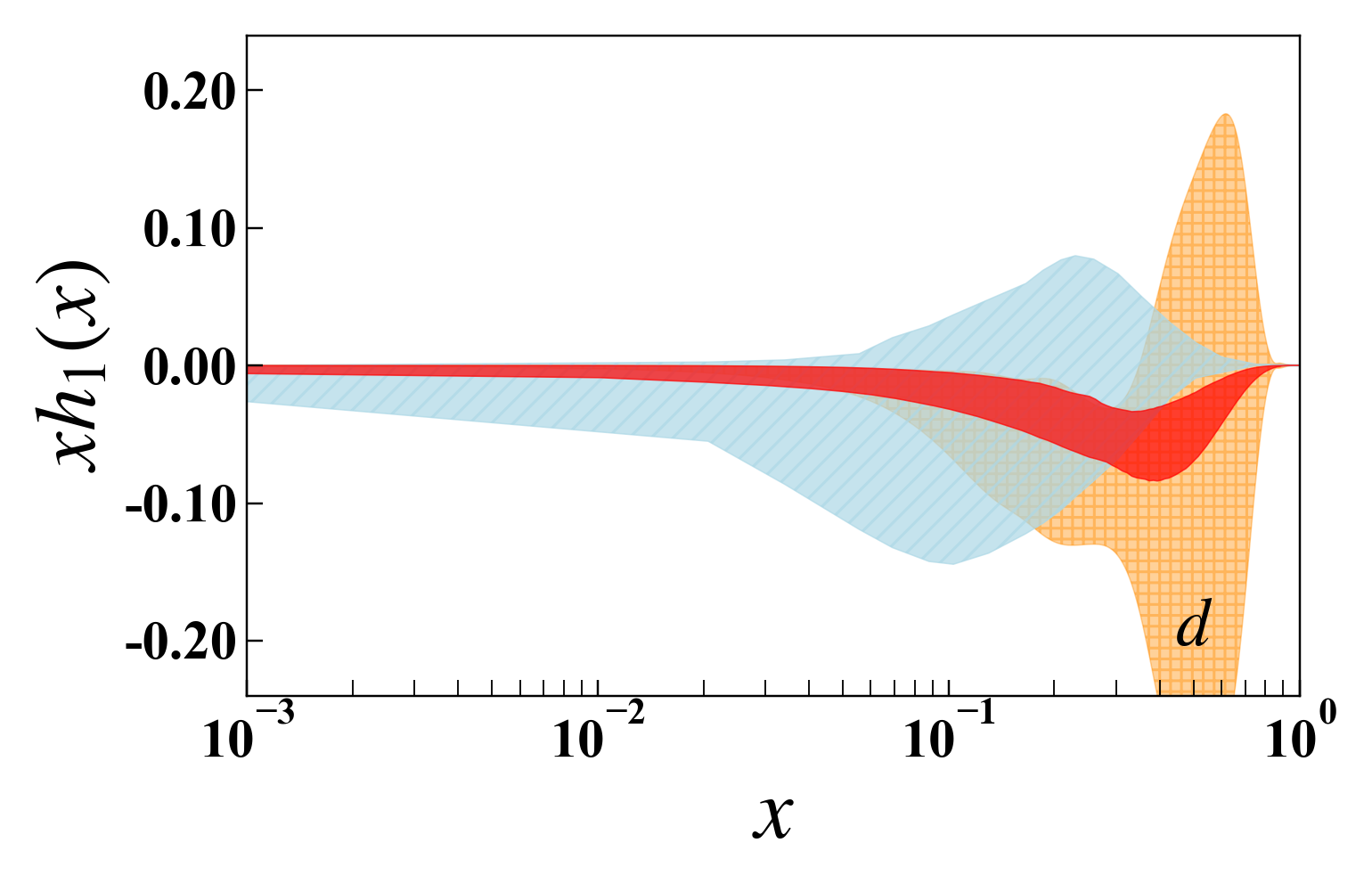}
    \caption{The extracted transverse momentum integrated transversity functions ($Q=2\, \rm GeV$). Our results are compared with Zeng et al.~\cite{Zeng:2023nnb} ($Q=2\, \rm GeV$) and Kang et al.~\cite{Kang:2015msa} ($Q=\sqrt{10}\, \rm GeV$). In this work, the transverse momentum integral as defined Eq.~(\ref{eq:xh1x}) is truncated at $k_T^{\text{cut}} = Q \times \delta_{\text{cut}}$ with $\delta_{\rm cut}=1$. In Zeng et al.~\cite{Zeng:2023nnb}, the integral is truncated at $k_T^{\rm cut} = 1\,\rm GeV$. }
\end{figure*}

\begin{table*}[htp]
\centering
\caption{The world SIDIS data used in our analysis with the $\chi^2/N$ of the
fit for Sivers and Collins asymmetries. $N$ stands for the number of data points, and the numbers in parentheses are the original
number of data points before applying the cut $\delta < 1$. }
\label{table:SIDIS_data}
\begin{tabular*}{0.9\textwidth}{m{0.15\textwidth}m{0.08\textwidth}m{0.1\textwidth}m{0.18\textwidth}m{0.1\textwidth}m{0.06\textwidth}m{0.06\textwidth}}
\hline\hline
Data set       & Target    & Beam       & Reaction &  $N$ & \multicolumn{2}{c}{$\chi^2/N$} \\ 
                &           &           &               &          & Sivers & Collins \\ \hline
HERMES~\cite{HERMES:2020ifk} &  $\text{H}_2$  &$27.6\,\rm GeV$ $e^{\pm}$    &$e^{\pm}p\to e^{\pm}\pi^+X $  &172 (192)   & 1.21  &   1.12  \\
& &&$e^{\pm}p\to e^{\pm}\pi^-X  $   &       &  &  \\
& &&$e^{\pm}p\to e^{\pm}K^+X  $         &    &  &  \\
& &&$e^{\pm}p\to e^{\pm}K^-X  $         &    &  &  \\
\hline
COMPASS~\cite{COMPASS:2008isr}  & $^{6}\text{LiD}$ & $160\,\rm GeV$ $\mu^+$   &$\mu^+d\to\mu^+\pi^+X $  &75 (104)  & 1.10 &  0.98\\                             & &&$\mu^+d\to\mu^+\pi^-X $        &   &  &\\
& &&$\mu^+d\to \mu^+K^+X  $          &   &  &\\
& &&$\mu^+d\to \mu^+K^-X  $     &       &  & \\
\hline
COMPASS~\cite{COMPASS:2014bze}  & $\text{NH}_3  $  &$160\,\rm GeV$ $\mu^+$  &$\mu^+p\to\mu^+\pi^+X $   &75 (104)    & 2.26 & 1.11\\
& &&$\mu^+p\to\mu^+\pi^-X $      &     &  &  \\
& &&$\mu^+p\to \mu^+K^+X  $      &       &  & \\
& &&$\mu^+p\to \mu^+K^-X  $      &     &  & \\
\hline
 COMPASS~\cite{ COMPASS:2023vhr}  & $^{6}\text{LiD}$ & $160\,\rm GeV$ $\mu^+$   &$\mu^+d\to\mu^+h^+X $  &38 (52)   & 0.83 & 1.07 \\                           & &&$\mu^+d\to\mu^+h^-X $       &   &  & \\
\hline
JLab2011~\cite{JeffersonLabHallA:2011ayy}  & $^{3}\text{He} $ & $5.9\,\rm GeV$ $e^-$  &$e^-n\to e^-\pi^+X $  &8 (8)   & 0.52 & 0.58 \\
& &&$e^-n\to e^-\pi^-X $        &     &  & \\
\hline
JLab2014~\cite{JeffersonLabHallA:2014yxb}  & $^{3}\text{He} $ & $5.9\,\rm GeV$ $e^-$  &$e^- {}^{3}{\rm He}\to e^-K^+X $  &5 (5)      & 1.22 & 1.48 \\
& &&$e^- {}^{3}{\rm He}\to e^-K^-X $        &      &  &  \\
\hline
  Total  &           &           &               &      373 (465)    & 1.35 & 1.08                                                      \\               
\hline \hline
\end{tabular*}
\end{table*}

\begin{table}[htp]
\centering
\caption{The Drell-Yan and $W^{\pm}/Z$-boson production data used in this analysis with the  $\chi^2/N$ of the fit results. $N$ stands for the number of data points, and the numbers in parentheses are the original
number of data points before applying the cut $\delta < 1$.}
\label{table:DY_data}
\begin{tabular*}{0.48\textwidth}{m{0.14\textwidth}m{0.15\textwidth}m{0.06\textwidth}m{0.08\textwidth}}
\hline\hline
Data set   &     Reaction & $N$ & $\chi^2/N$ \\
\hline
COMPASS~\cite{COMPASS:2023vqt}  &    $\pi^- + p^{\uparrow}\to\gamma^*+X $  & 15 (15) & 0.79 \\ 
  STAR.W+~\cite{STAR:2015vmv}  &  $p^{\uparrow}+p\to W^++X$& 8 (8) &2.23\\
  STAR.W-~\cite{STAR:2015vmv} &  $p^{\uparrow}+p\to W^-+X$& 8 (8) &1.76\\
  STAR.Z~\cite{Collaboration:2023oml}& $p^{\uparrow}+p\to \gamma^*/Z+X$& 1 (1)& 0.48\\
\hline
Total&  &32 (32) & 1.38 \\
\hline \hline
\end{tabular*}
\end{table}

\begin{table}[htp]
\centering
\caption{The world SIA data used in this analysis  with the $\chi^2/N$ of the fit result. $N$ stands for the number of  data points, and the numbers in parentheses are the original
number of data points before applying the cut $\delta < 1$.}
\label{table:SIA_data}
\begin{tabular*}{0.48\textwidth}{m{0.14\textwidth}m{0.13\textwidth}m{0.08\textwidth}m{0.08\textwidth}}
\hline\hline
Data set                   & Reaction              &$N$ & $\chi^2/N$\\ \hline
BELLE~\cite{Belle:2008fdv}  &$e^+e^-\to\pi\pi X $   & 16 (16) & 0.79 \\
BABAR~\cite{BaBar:2013jdt}   &$e^+e^-\to\pi\pi X $   & 45 (45) &1.04\\
BABAR~\cite{BaBar:2015mcn}   &$e^+e^-\to\pi\pi X $    & 48 (48)  & 0.79   \\
&$e^+e^-\to\pi K X $      &      &   \\
&$e^+e^-\to K K X $       &      & \\
BESIII~\cite{BESIII:2015fyw} &$e^+e^-\to\pi\pi X $     & 11 (11)& 2.24\\
\hline
Total&  &120 (120) & 1.01 \\
\hline \hline
\end{tabular*}
\end{table}

\section{Summary}\label{sec:summary}

We have carried out a phenomenological investigation, aiming at extracting the Sivers functions, transversity distribution functions, and Collins fragmentation functions. This global analysis includes a comprehensive fitting procedure that combines the latest COMPASS and STAR data from SIDIS, Drell-Yan, and $W^{\pm}/Z$-boson production processes, along with pre-existing data from SIDIS and SIA measurements. Both statistical and systematic uncertainties are taken into account, including the correlated relative scale uncertainties, in the calculation of the $\chi^2$ and in the generation of replicas.
Due to the limited amount of DY and $W^{\pm}/Z$-boson production data, the improvement by the addition of these data is marginal on the determination of TMD functions, and the extracted of TMD distributions remain predominantly governed by the SIDIS data.  
However, a notable improvement is observed by incorporating the latest SIDIS data on transversely polarized deuteron targets from the COMPASS experiment~\cite{ COMPASS:2023vhr}. It has significantly reduced the uncertainties of the distributions of $d$ and $\bar{d}$ quarks for both transversity and Sivers functions, as well as the tensor charge.

\section*{Acknowledgments}
The authors acknowledge the computing resource at the Southern Nuclear Science Computing Center.
T.L. is supported by the National Natural Science Foundation of China under grants No.~12175117 and No.~12321005, and
Shandong Province Natural Science Foundation Grant
No. ZFJH202303.
P.S. is supported by the Natural Science Foundation of China under grants No.~11975127 and No.~12061131006.
Y.Z. is supported by the Strategic Priority Research Program of the Chinese Academy of Sciences under grant No.~XDB34000000, the Guangdong Major Project of Basic and Applied Basic Research No.~2020B0301030008, and CAS Project for Young Scientists in Basic Research No.YSBR-117.

\appendix

\section{Creating of  world data replicas}
 \label{App:smear_replicas}
 The $n$ data points with correlated overall relative scale uncertainty $\sigma^{cor.}$ can be described by a data set as 
 \begin{align}
     y=\{m_1, m_2, \dots, m_i, \dots, m_n \},
 \end{align} 
where $m_i$ is the central value of the each data point. Then for each replica, the center value correction is defined as
\begin{align}
    c_i^{rep.}=\mathrm{Random}(0, \sigma^{cor.}),
\end{align}
where $\mathrm{Random}(m, \sigma)$ is a function to return a random number of a Gaussian distribution with center value $m$ and Gaussian width $\sigma$. Then a replica of $n$ world data points can be written as
\begin{align}
     y^{rep.}=\{m^{rep.}_1, m^{rep.}_2, \dots, m^{rep.}_i, \dots, m^{rep.}_n \},
\end{align}
with
\begin{align}
    m_i^{rep.}=m_i^{ran.} +c_i^{rep.} m_i^{ran.},
\end{align}
where
\begin{align}
    m_i^{ran.}=\mathrm{Random}(m_i, \sigma_i^{uncor.}).
\end{align}
Here $\sigma_i^{uncor.}$ is the uncorrelated uncertainty of $i'$th measurement.

\bibliographystyle{elsarticle-num}







\end{document}





\title{Supplemental Material: comparisons with experimental data}

\maketitle

\begin{figure*}[htp]
    \centering
    \includegraphics[width=0.48\textwidth]{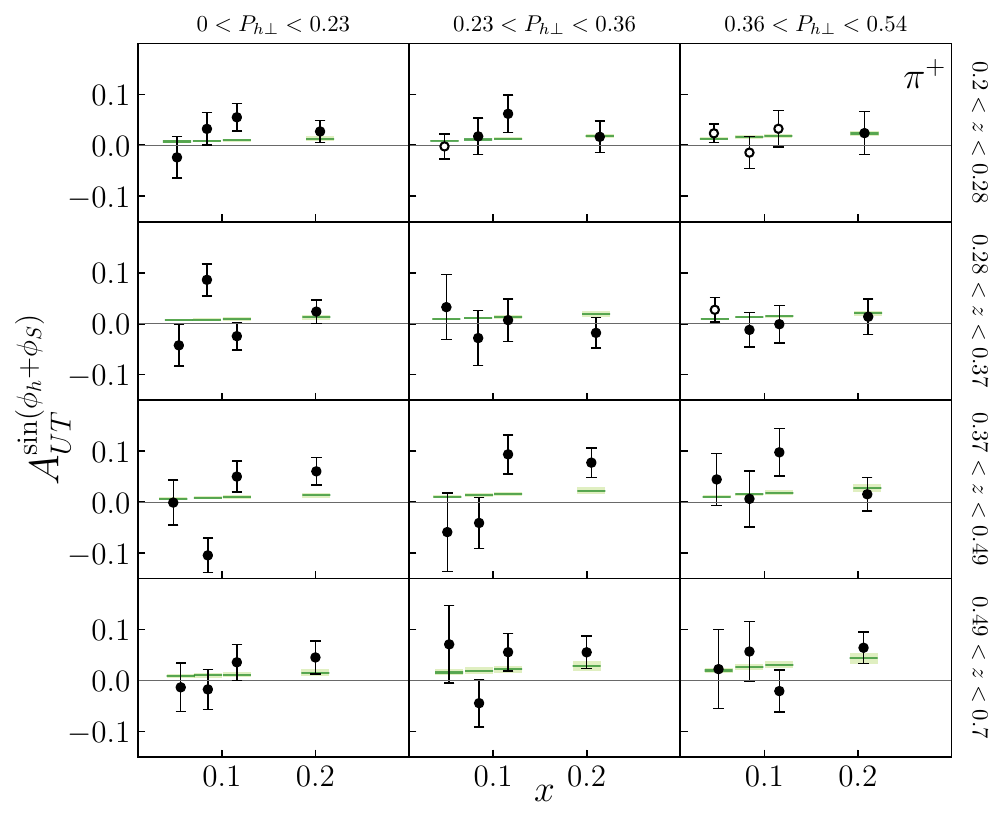}
    \includegraphics[width=0.48\textwidth]{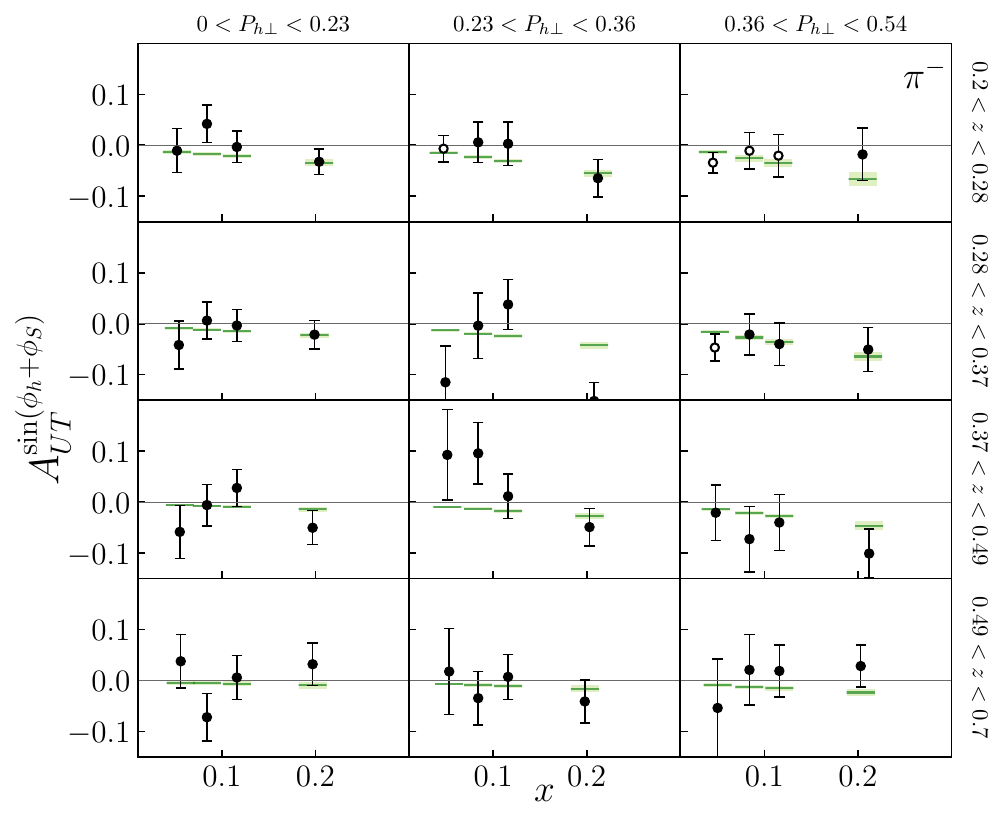}
    \includegraphics[width=0.48\textwidth]{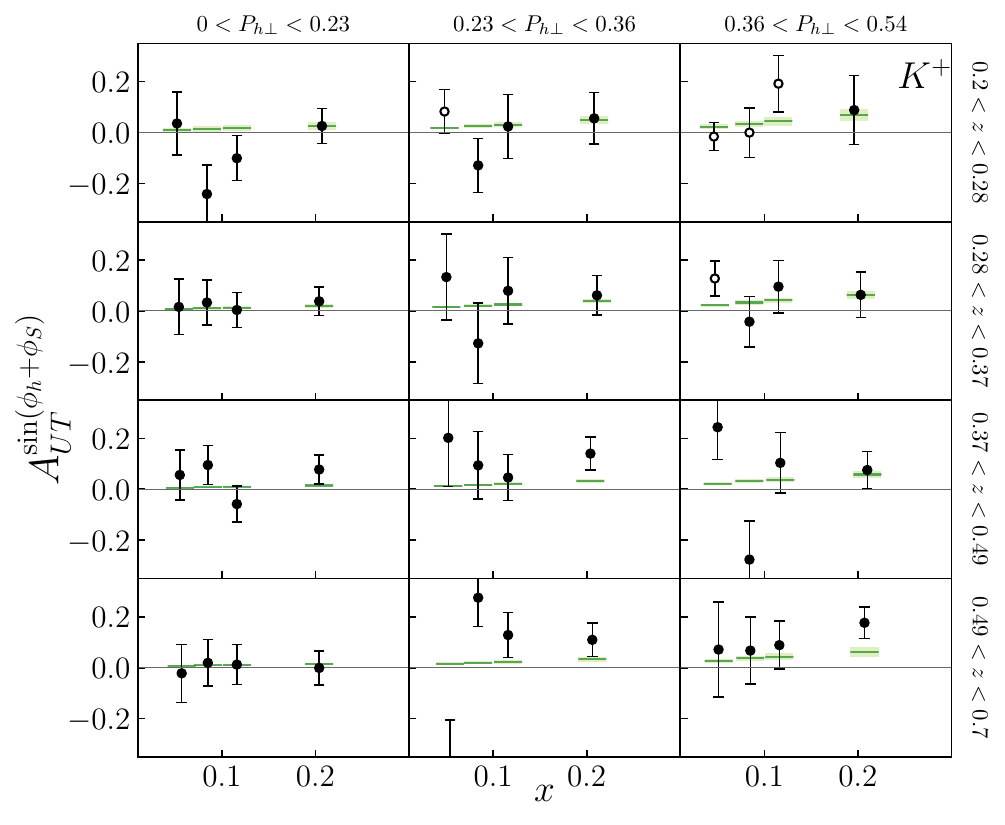}
    \includegraphics[width=0.48\textwidth]{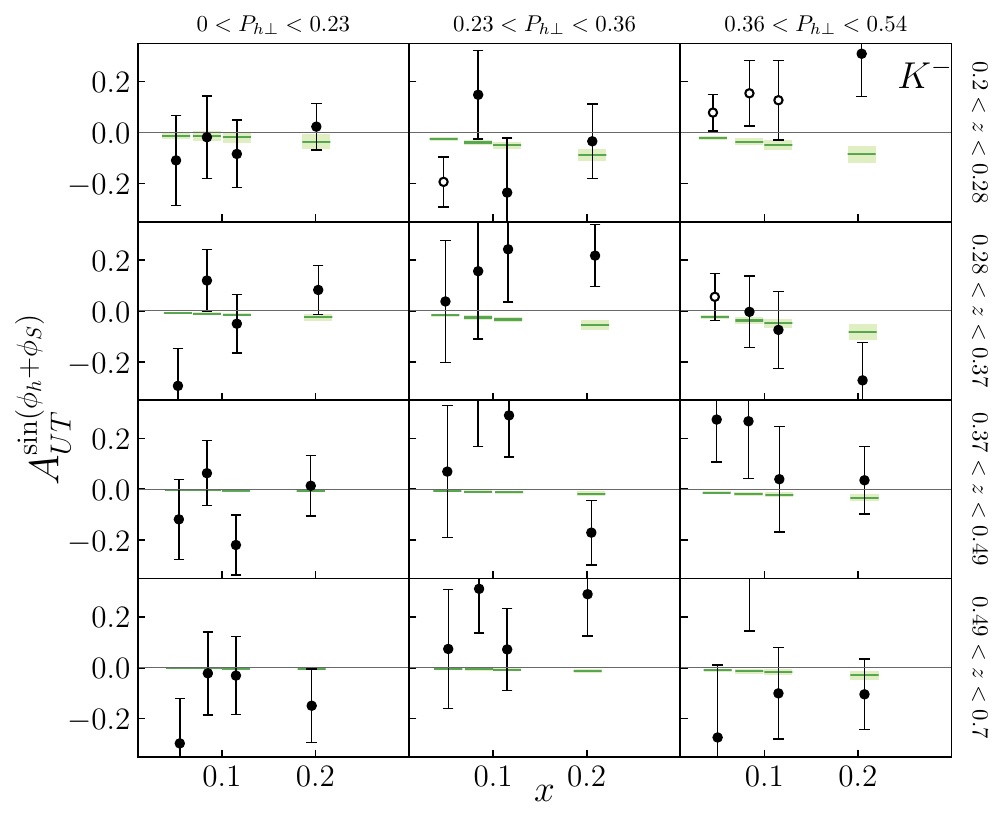}
    \caption{Comparison with HERMES Collins SSA data~\cite{HERMES:2020ifk} from the proton target for $\pi^+$ (upper left), $\pi^-$ (upper right), $K^+$ (lower left), and $K^-$ (lower right) productions. }
    \label{fig:collins_hermes}
\end{figure*}
\begin{figure*}[htp]
    \centering
    \includegraphics[width=0.48\textwidth]{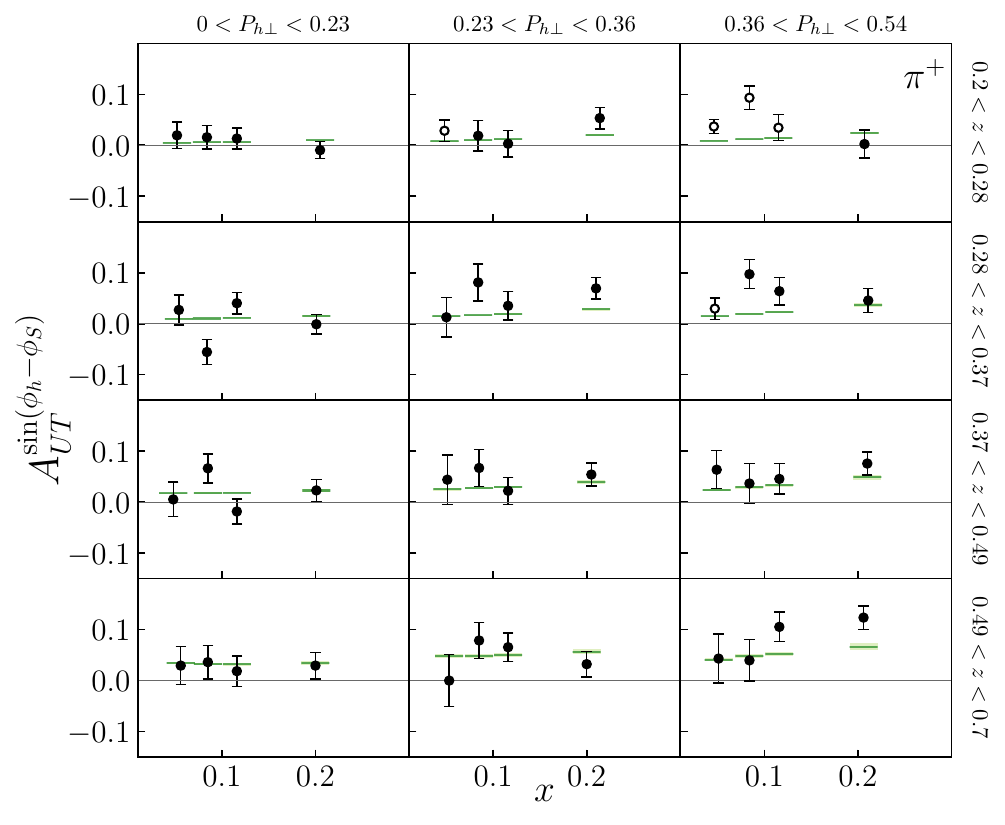}
    \includegraphics[width=0.48\textwidth]{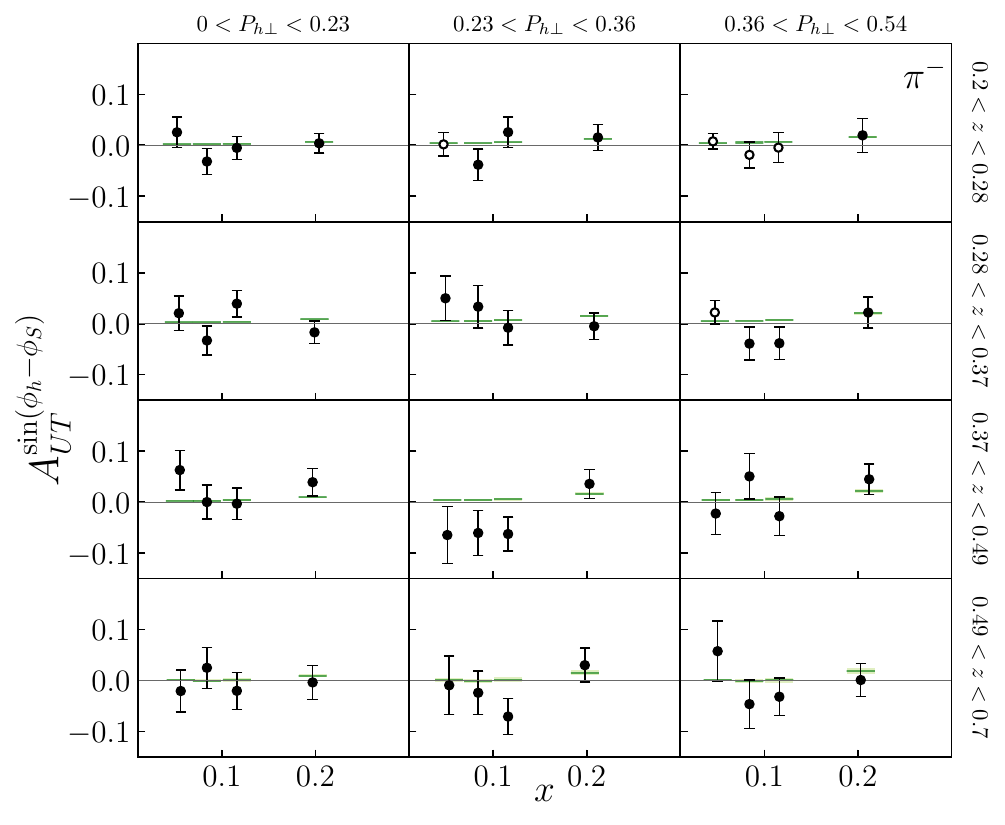}
    \includegraphics[width=0.48\textwidth]{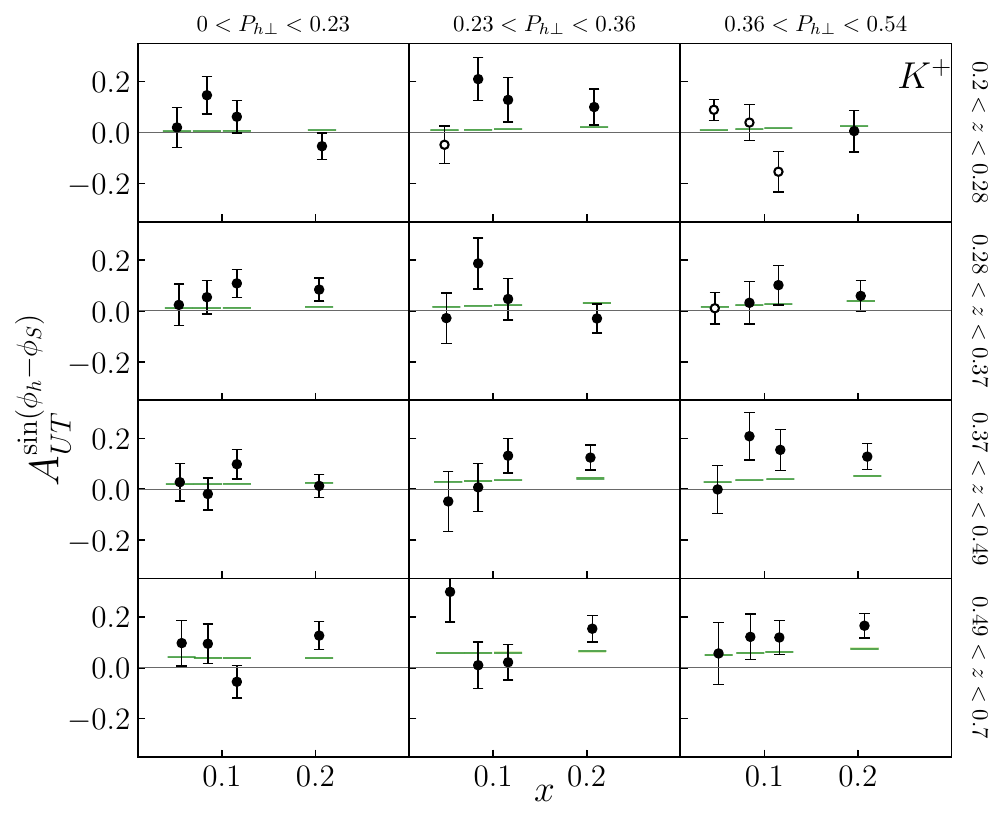}
    \includegraphics[width=0.48\textwidth]{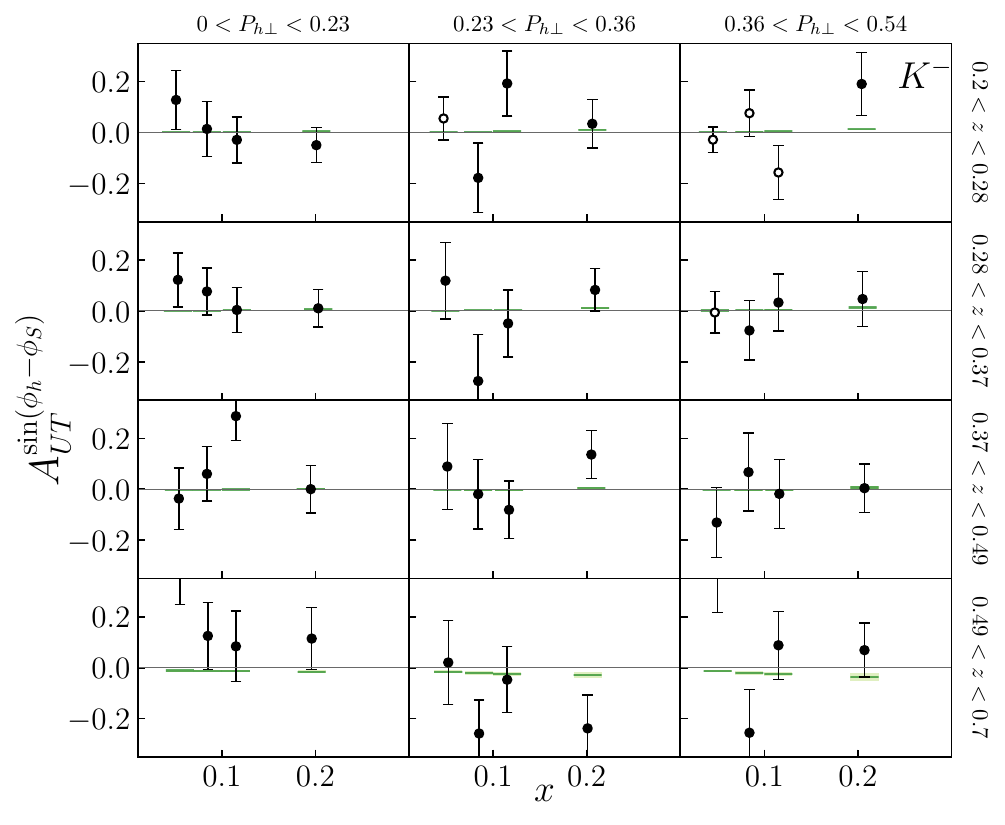}
    \caption{Comparison with HERMES Sivers SSA data~\cite{HERMES:2020ifk} from the proton target for $\pi^+$ (upper left), $\pi^-$ (upper right), $K^+$ (lower left), and $K^-$ (lower right) productions.}
    \label{fig:sivers_hermes}
\end{figure*}
\begin{figure*}[htp]
    \centering
    \includegraphics[width=0.48\textwidth]{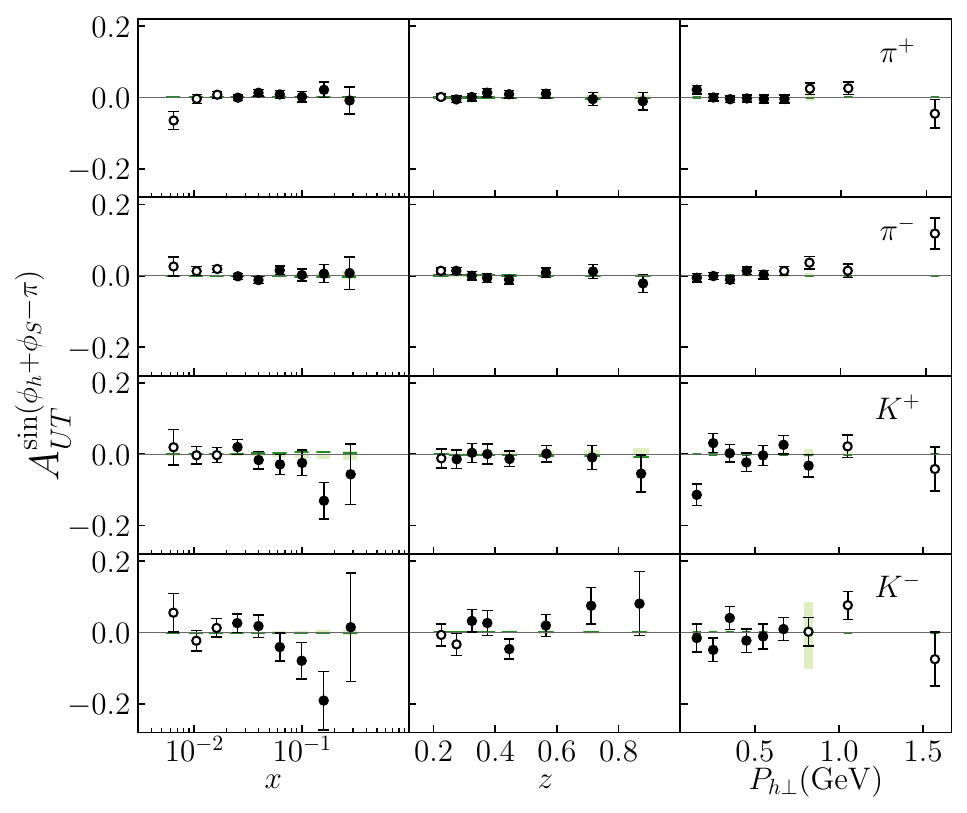}
    \includegraphics[width=0.48\textwidth]{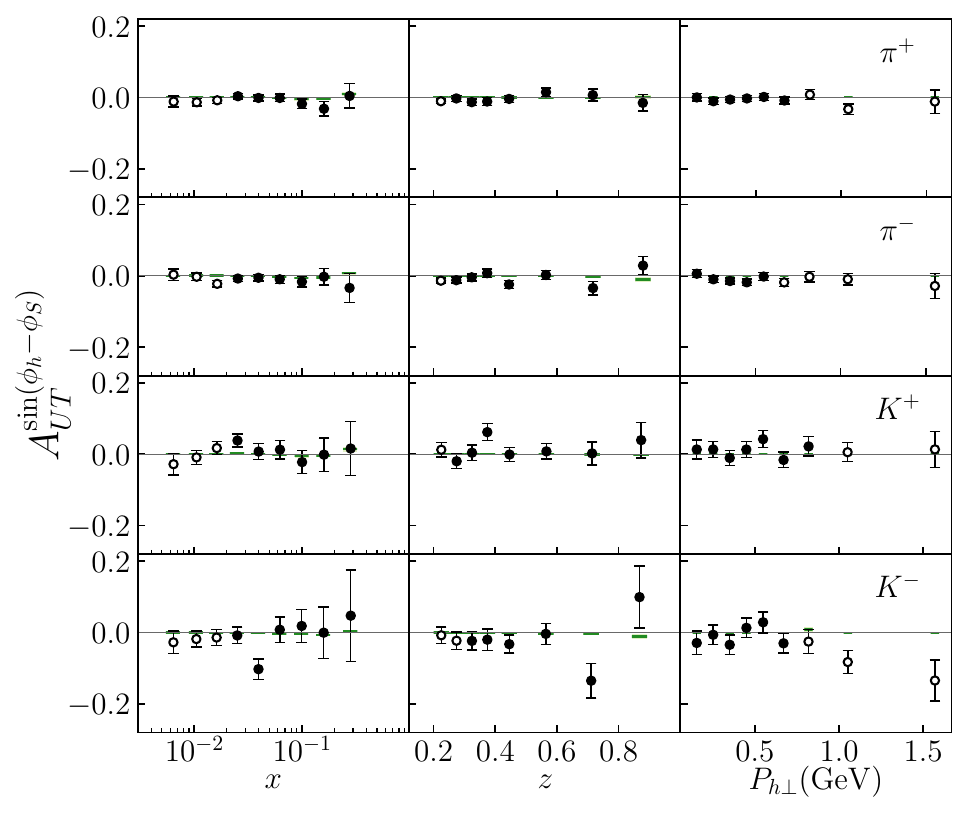}
    \caption{Comparison with COMPASS Collins (left) and Sivers (right) SSA data~\cite{COMPASS:2008isr} from the deuteron target for $\pi^+$, $\pi^-$, $K^+$, and $K^-$ productions.}
    \label{fig:compass_deuteron}
\end{figure*}
\begin{figure*}[htp]
    \centering
    \includegraphics[width=0.48\textwidth]{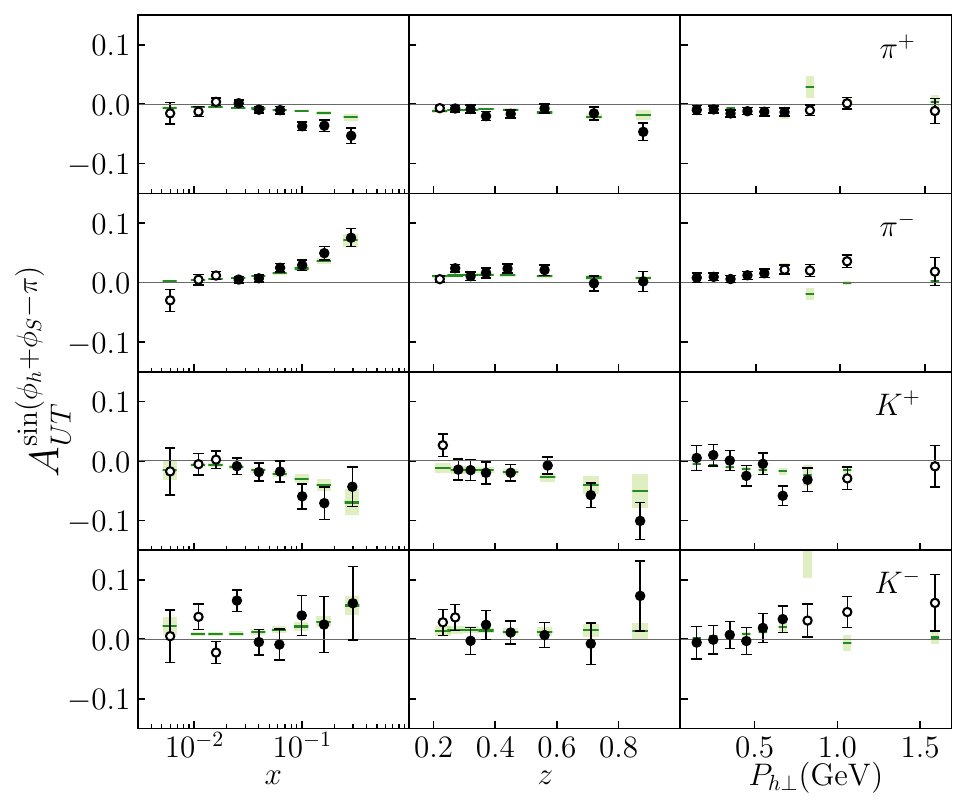}
    \includegraphics[width=0.48\textwidth]{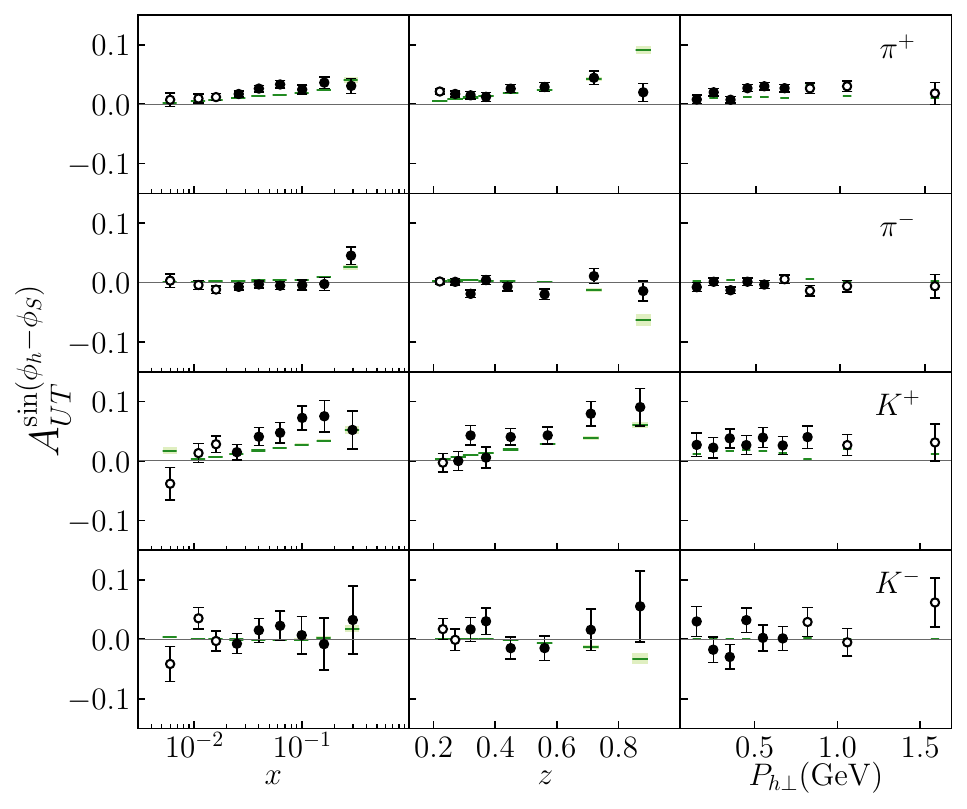}
    \caption{Comparison with COMPASS Collins (left) and Sivers (right) SSA data~\cite{COMPASS:2014bze} from the proton target for $\pi^+$, $\pi^-$, $K^+$, and $K^-$ productions.}
    \label{fig:compass_proton}
\end{figure*}
\begin{figure*}[htp]
    \centering
    \includegraphics[width=0.48\textwidth]{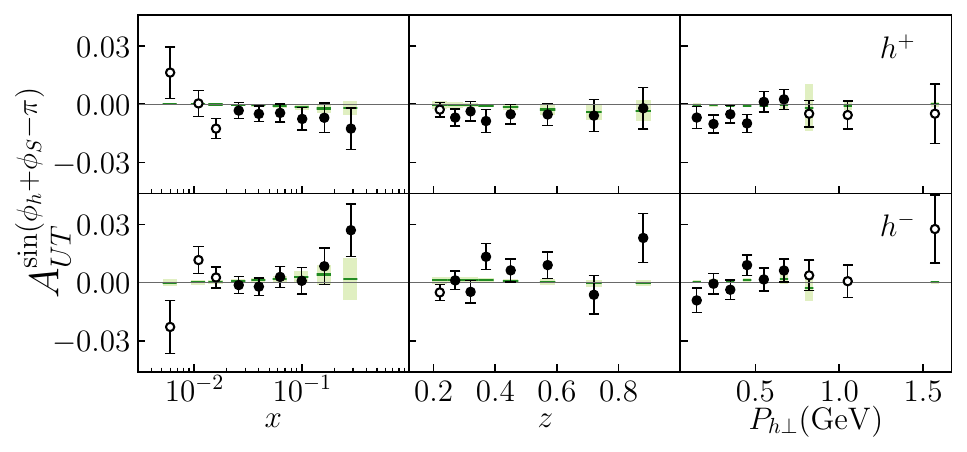}
    \includegraphics[width=0.48\textwidth]{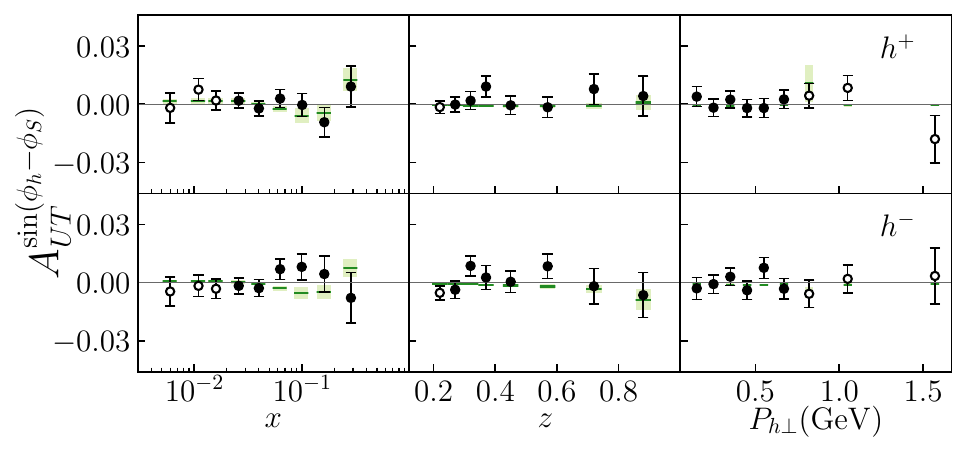}
    \caption{Comparison with COMPASS Collins (left) and Sivers (right) SSA data~\cite{COMPASS:2023vhr} from the deuteron target for $h^+$, and $h^-$ productions.}
    \label{fig:compass2022_deuteron}
\end{figure*}
\begin{figure*}[htp]
    \centering
    \includegraphics[width=0.48\textwidth]{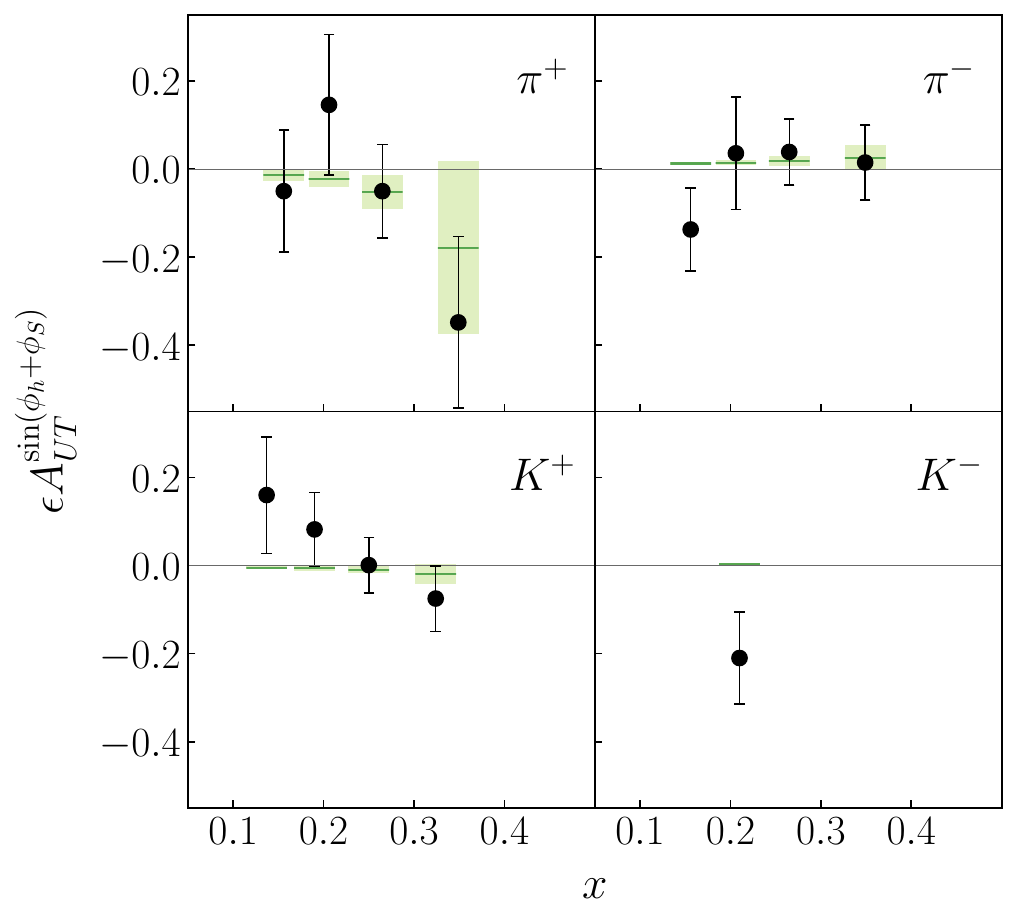}
    \includegraphics[width=0.48\textwidth]{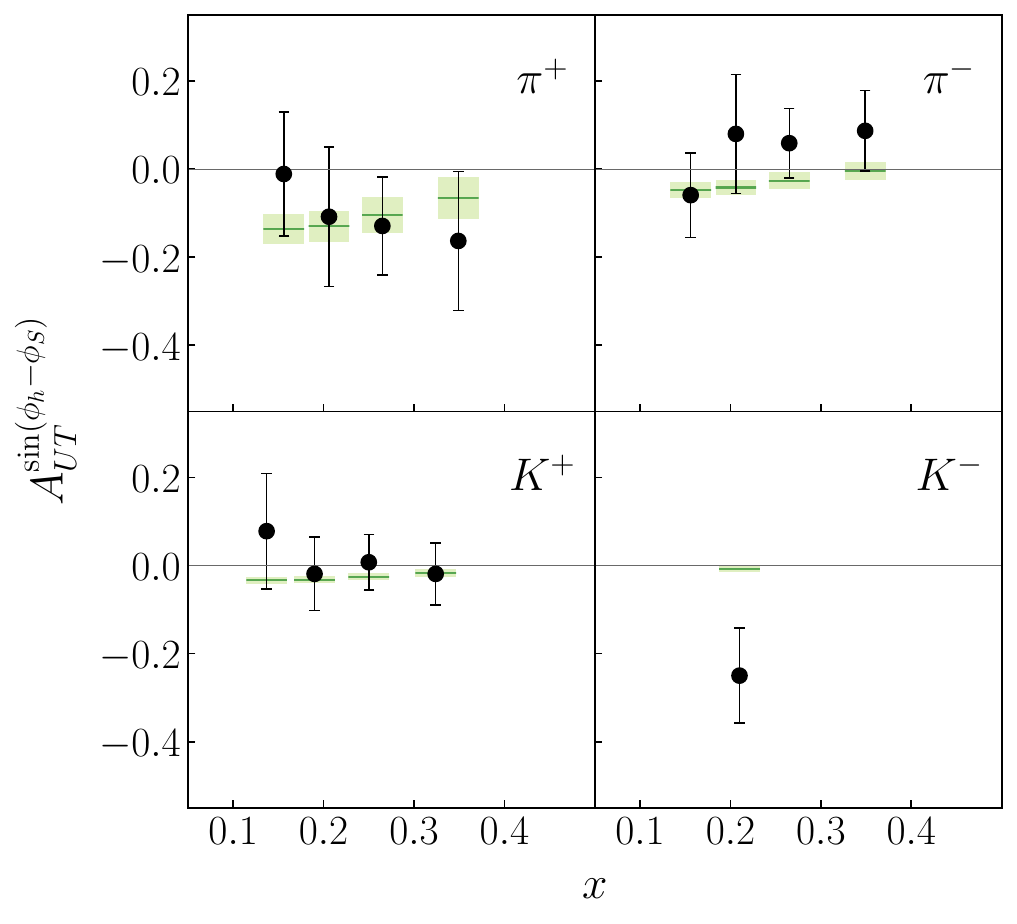}
    \caption{Comparison with JLab Collins (left) and Sivers (right) SSA data~\cite{JeffersonLabHallA:2011ayy,JeffersonLabHallA:2014yxb} from the $^3\rm He$ target. The asymmetry values for $\pi^+$ and $\pi^-$ productions  have been effectively converted to those of the neutron, and the asymmetry values for $K^+$ and $K^-$ productions. }
    \label{fig:jlab}
\end{figure*}
\begin{figure*}[htp]
    \centering
    \includegraphics[width=0.6\textwidth]{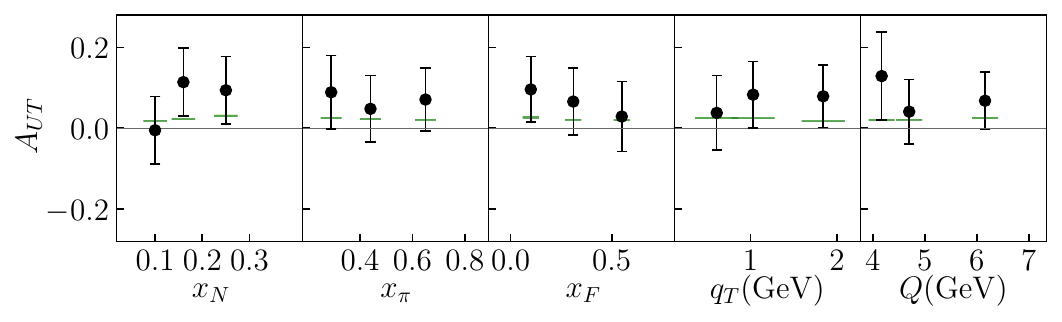}
    \caption{Comparison with COMPASS DY data~\cite{COMPASS:2023vqt}.}
    \label{fig:compassDY}
\end{figure*}
\begin{figure*}[htp]
    \centering
    \includegraphics[width=0.35\textwidth]{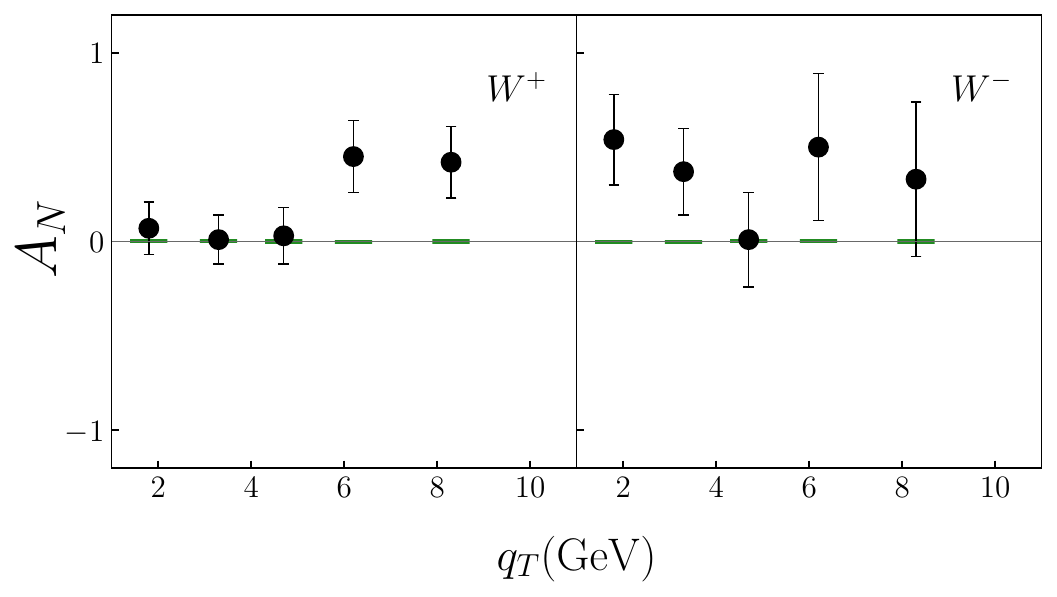}
    \includegraphics[width=0.49\textwidth]{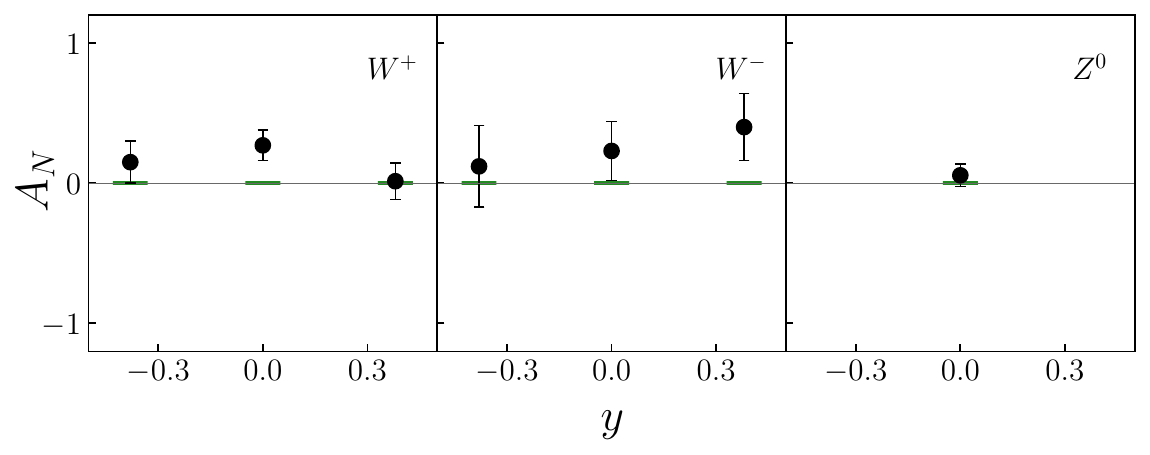}
    \caption{Comparison with STAR Sivers SSA data~\cite{STAR:2015vmv, Collaboration:2023oml} for $W^{\pm}$ and $Z$ boson production.}
    \label{fig:STAR}
\end{figure*}
\begin{figure}[htp]
    \centering
    \includegraphics[width=0.6\textwidth]{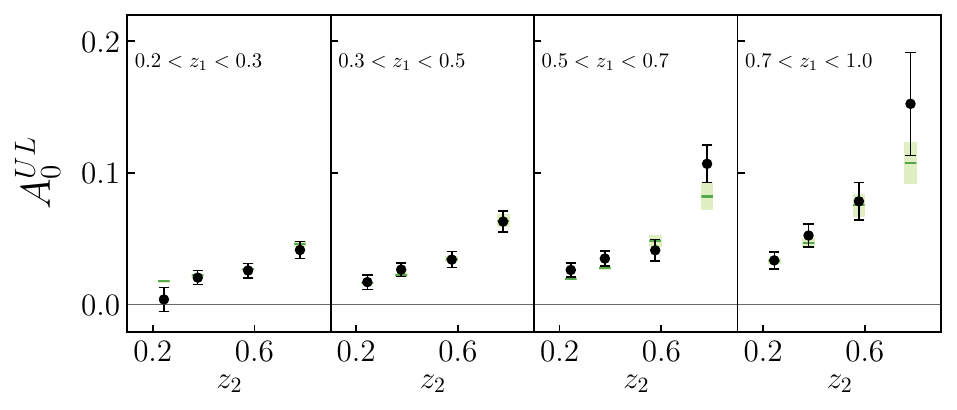}
    \caption{Comparison of BELLE $\pi\pi$ channel Collins asymmetry data~\cite{Belle:2008fdv} to theoretical calculations in the SIA process.}
    \label{fig:Belle}
\end{figure}
\begin{figure*}[htp]
    \centering
    \includegraphics[width=0.4\textwidth]{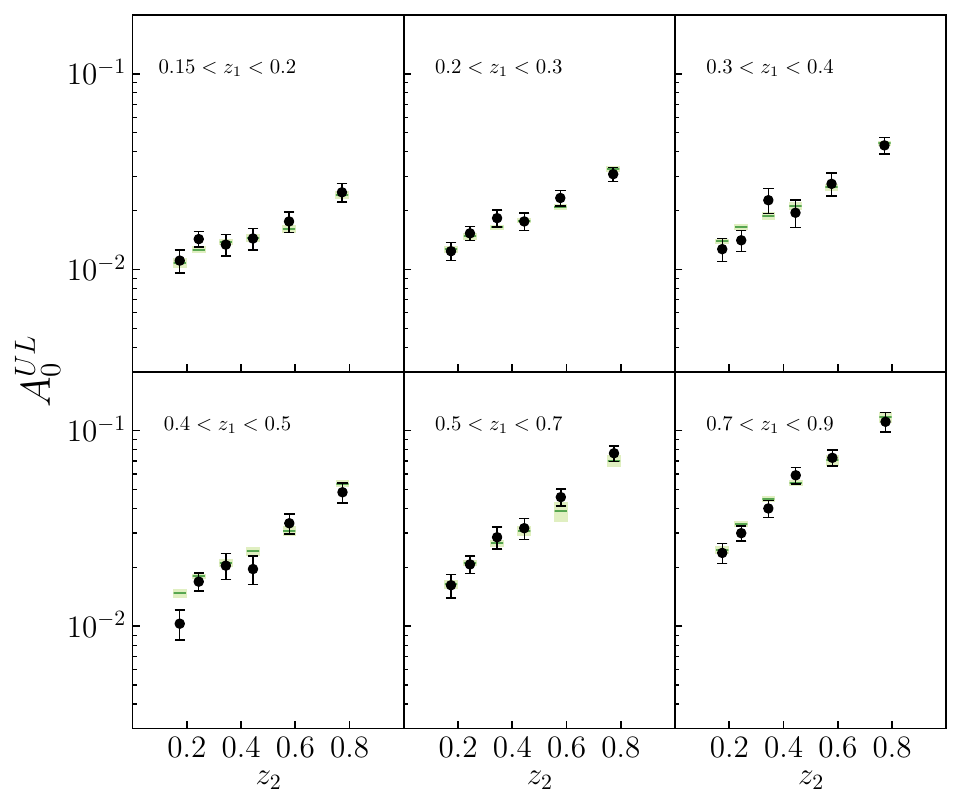}
    \includegraphics[width=0.4\textwidth]{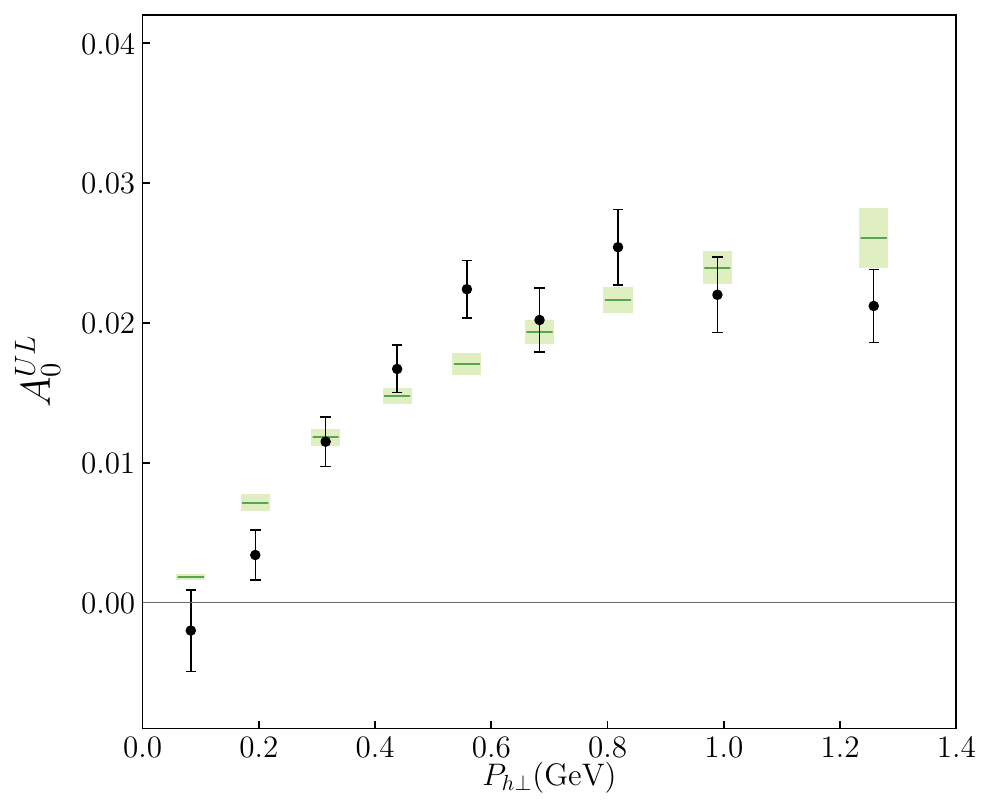}
    \caption{Comparison of BABAR $\pi\pi$ channel Collins asymmetry data~\cite{BaBar:2013jdt} to theoretical calculations in the SIA process as a function of $z$ (left) and $P_{h\perp}$ (right).}
    \label{fig:Babar2014}
\end{figure*}
\begin{figure}[htp]
    \centering
    \includegraphics[width=0.4\textwidth]{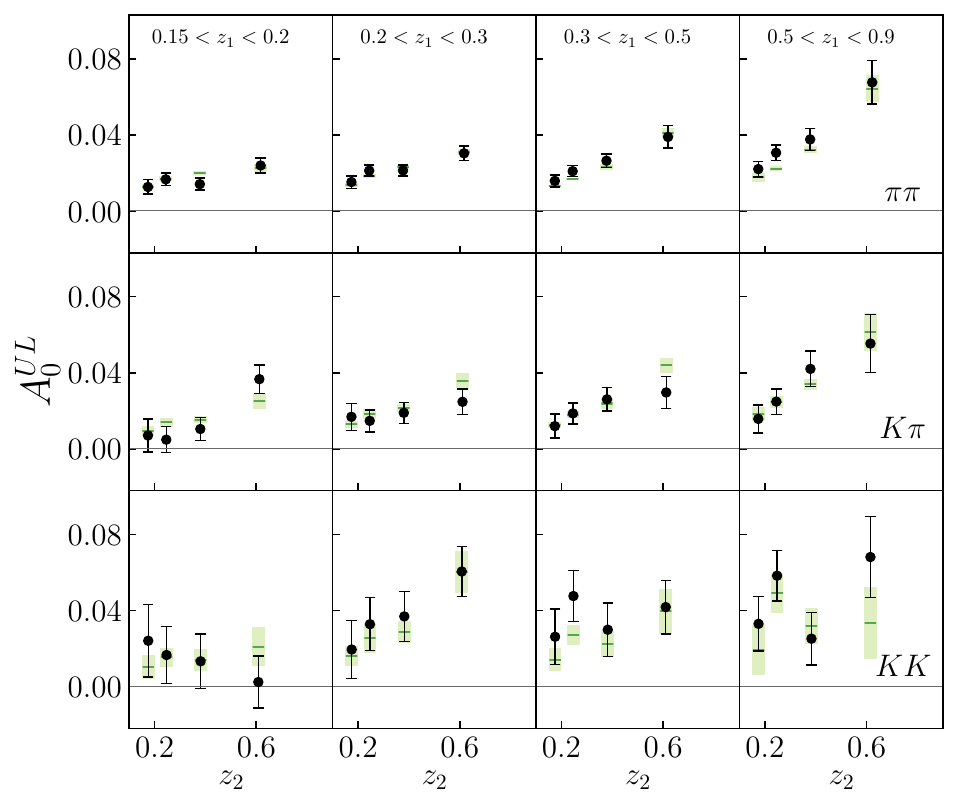}
    \caption{Comparison of BABAR $\pi\pi$, $K\pi$ and $KK$ channels Collins asymmetry data~\cite{BaBar:2015mcn} to theoretical calculations in the SIA process.}
    \label{fig:Babar2016}
\end{figure}
\begin{figure}[htp]
    \centering
    \includegraphics[width=0.4\textwidth]{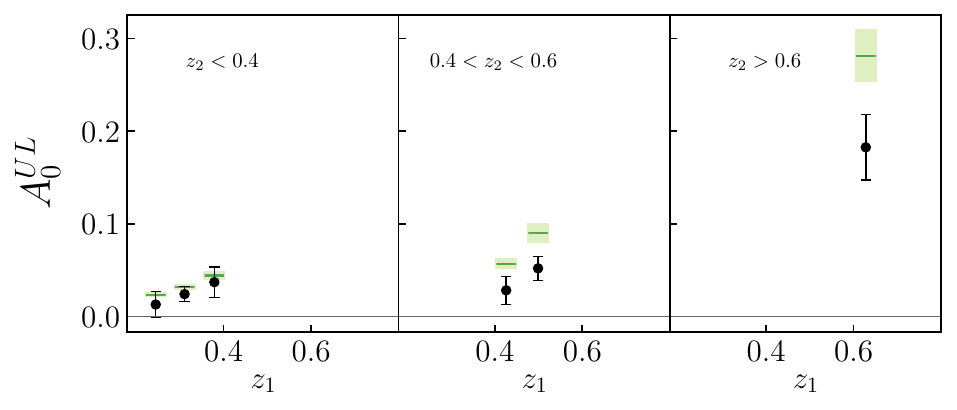}
    \includegraphics[width=0.4\textwidth]{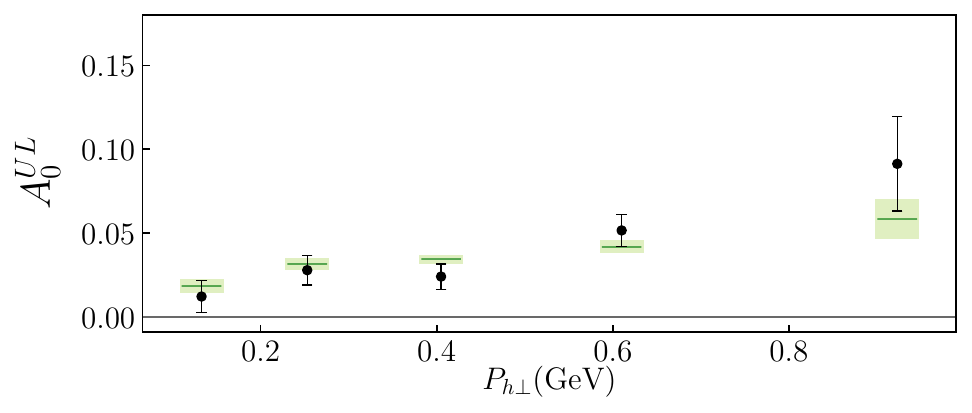}
    \caption{Comparison of BESIII $\pi\pi$ channel Collins asymmetry data~\cite{BESIII:2015fyw} to theoretical calculations in the SIA process as a function of $z$ (left) and $P_{h\perp}$ (right).}
    \label{fig:BESIII}
\end{figure}

The comparisons between experimental data and the calculations by using replicas are shown in Figs.~\ref{fig:collins_hermes} to~\ref{fig:BESIII}, The filled data points are included in the fit, while the open data points are not. The green lines are the central value calculated from the fit and the bands represent the standard deviation calculated from 1000 replicas.

